\documentclass[preprint]{aastex631}

\newcommand\kpc{\rm kpc}

\usepackage{CJK}
\usepackage[para]{threeparttable}
\usepackage{xcolor}

\begin{document}
\begin{CJK*}{UTF8}{gbsn}

\title[The Effect of Galaxy Interactions]{The Effect of Galaxy Interactions on Starbursts in Milky Way-Mass Galaxies in FIRE Simulations}

\correspondingauthor{Fei Li}
\email{lif.li@mail.utoronto.ca}

\author[0000-0002-1666-7067]{Fei Li (李菲)}
\affiliation{David A. Dunlap Department of Astronomy and Astrophysics, University of Toronto, 50 St. George Street, ON M5S 3H4, Canada.}
\affiliation{Canadian Institute for Theoretical Astrophysics, University of Toronto, 60 St. George Street, Toronto, ON M5S 3H8, Canada.}

\author[0000-0003-1842-6096]{Mubdi Rahman}
\affiliation{Sidrat Research, 124 Merton Street, Suite 507, Toronto, ON M4S 2Z2, Canada}

%\collaboration{20}{(AAS Journals Data Editors)}

\author[0000-0002-8659-3729]{Norman Murray}
\affiliation{Canadian Institute for Theoretical Astrophysics, University of Toronto, 60 St. George Street, Toronto, ON M5S 3H8, Canada.}
\affiliation{Canada Research Chair in Astrophysics.}

\author[0000-0002-1666-7067]{Du{\v s}an Kere{\v s}}
\affiliation{Department of Physics, Center for Astrophysics and Space Sciences, University of California at San Diego, 9500 Gilman Drive, \\ 
La Jolla, CA 92093, USA.}

\author[0000-0003-0603-8942]{Andrew Wetzel}
\affiliation{Department of Physics \& Astronomy, University of California, Davis, CA 95616, USA.}

\author[0000-0002-4900-6628]{Claude-Andr{\'e} Faucher-Gigu{\`e}re}
\affiliation{Department of Physics and Astronomy and CIERA, Northwestern University, 2145 Sheridan Road, Evanston, IL 60208, USA.}

\author[0000-0003-3729-1684]{Philip F. Hopkins}
\affiliation{TAPIR, Mailcode 350-17, California Institute of Technology, Pasadena, CA 91125, USA.}

\author[0000-0002-3430-3232]{Jorge Moreno}
\affiliation{Department of Physics and Astronomy, Pomona College, Claremont, CA 91711, USA}
\affiliation{The Observatories of the Carnegie Institution for Science, 813 Santa Barbara Street, Pasadena, CA 91101, USA}


\begin{abstract} %250 word limit for ApJ

Simulations and observations suggest that galaxy interactions may enhance the star formation rate (SFR) in merging galaxies. One proposed mechanism is the torque exerted on the gas and stars in the larger galaxy by the smaller galaxy. We analyze the interaction torques and star formation activity on six galaxies from the FIRE-2 simulation suite with masses comparable to the Milky Way galaxy at redshift $z=0$. We trace the halos from $z = 3.6$ to $z=0$, calculating the torque exerted by the nearby galaxies on the gas in the central galaxy. We calculate the correlation between the torque and the SFR across the simulations for various mass ratios. For near-equal-stellar-mass-ratio interactions in the galaxy sample, occurring between $z=1.2-3.6$, there is a positive and statistically significant correlation between the torque from nearby galaxies on the gas of the central galaxies and the SFR. For all other samples, no statistically significant correlation is found between the torque and the SFR. Our analysis shows that some, but not all, major interactions cause starbursts in the simulated Milky Way-mass galaxies, and that most starbursts are not caused by galaxy interactions. The transition from `bursty' at high redshift ($z\gtrsim1$) to `steady' star-formation state at later times is independent of the interaction history of the galaxies, and most of the interactions do not leave significant imprints on the overall trend of the star formation history of the galaxies.
\end{abstract}



\section{Introduction}

Simulations of galaxies that resolve the interstellar medium (ISM) at giant molecular cloud-scales show that massive galaxies transition from a ``bursty'' star formation state with large temporal fluctuations in star formation rate (SFR) at high redshift to a ``steady'' star formation state at low redshift \citep[][]{Hopkins2014, Muratov2015MNRAS.454.2691M, Sparre2017MNRAS.466...88S, Hopkins2018, Faucher2018MNRAS.473.3717F}. It is worth noting that this connects broadly with the transition from a thick to thin disk geometry \citep{Yu2021MNRAS.505..889Y, Yu2023MNRAS.523.6220Y,  McCluskey2024MNRAS.527.6926M}. The physical cause of this transition, however, remains uncertain \citep{Stern2021ApJ...911...88S, Gurvich2023MNRAS.519.2598G, Hopkins2023}. A proposed mechanism is the change in merger and flyby frequency with redshift \citep{Barnes1996}. In this paper, we investigate this mechanism by looking at the effects of galaxy interactions on the star formation activity of central galaxies and their likelihood to trigger starburst events.

Mergers and flybys \citep{Moreno2012MNRAS.419..411M, Moreno2013MNRAS.436.1765M} are common in the growth history of galaxies in the $\Lambda$CDM cosmological model, and it is commonly believed that galaxy interactions can induce star formation in central galaxies. Previous simulations have shown that the torque from a companion galaxy exerted on the central galaxy can lead to the inflow of gas, and thus to the enhancement of the SFR in the central galaxy \citep{Toomre1972, Keel1985, Sanders1988ApJ...325...74S, Barnes1996, Mihos1996, Tissera2002, Cox2006, Montuori2010, Rupke2010, Torrey2012}. Consistent with the insight from the simulations, there have also been observations \citep{Larson1978, Barton2000, Alonso2004, Nikolic2004, Woods2006, Woods2007, Barton2007, Ellison2008, Heiderman2009, Knapen2009, Robaina2009, Ellison2010, Woods2010, Patton2011} showing that, near redshift zero, the presence of nearby galaxies coincides with higher SFR, and that the separation of companions is inversely correlated with the SFR.

Compared to galaxies in the star-forming main sequence, a larger fraction of starburst galaxies display merger features \citep{Luo2014ApJ...789L..16L, Knapen2015_starburst, Blumenthal2020MNRAS.492.2075B}. In observations of the local universe, the majority of ultraluminous infrared galaxies (ULIRGs; $L_{IR} \ge 10^{12} L_{\odot}$, i.e., extreme starbursts), display features of strongly interacting systems such as bridges, tidal tails, and disturbed morphologies \citep[e.g.,][]{Veilleux2002ApJS..143..315V, Colina2005ApJ...621..725C, Gao2004ApJ...606..271G, Garcia2012A&A...539A...8G, Hung2013ApJ...778..129H}.

At higher redshift, mergers are more common, and if the impact of mergers on the SFR activity is similar to the local universe, it is reasonable to assume that a larger fraction of the stellar mass created in that epoch is directly attributable to mergers. However, whether the effect of galaxy interactions on SFR enhancement at high redshift is different from the local universe is unclear. Some observational studies \citep{Rodighiero2011, Schreiber2015, Lofthouse2017, Silva2018, Pearson2019} show a measurable but marginally statistically significant enhancement of the SFR in merging galaxies compared to the non-merging sample. There are simulations of idealized galaxies \citep{Perret2014, Scudder2015, Fensch2017} with gas fractions typical of high-redshift galaxies, which found that the merger-induced star formation enhancement is weaker compared to galaxies with lower gas fractions.

As the tidal features of galaxy interactions with lower mass ratios are more difficult to observe (especially at high redshift) due to lower surface brightness, minor interactions \citep[stellar mass ratios $\mu$ $\le$ 0.25;][]{Jackson2022MNRAS.511..607J} and mini interactions \citep[$\mu$ $\le$ 0.1;][]{Bottrell2024MNRAS.527.6506B} have been difficult to observe, but new facilities may improve this situation. Galaxy simulations that can predict the observable properties can be useful in studying the effect of galaxy interactions of various mass ratios on the star formation history. 

The FIRE\footnote{\url{http://fire.northwestern.edu}} suite of simulations \citep{Hopkins2014, Hopkins2018} provide a useful laboratory to investigate how interactions affect the internal star formation properties of a galaxy. These simulations explicitly model the radiation pressure, stellar winds, and ionization of young stellar populations to account for the effects of feedback from star formation on their natal environment. This provides a physically realistic model of star formation that has shown success in reproducing a diverse range of observational results, from the relationship between stellar mass and halo mass of galaxies \citep{Hopkins2014, Feldmann2017MNRAS.470.1050F, Hopkins2018} to the mass-metallicity relationship of stars and gas within galaxies \citep{2016MNRAS.456.2140M, Wetzel2016, Bassini2024MNRAS.532L..14B, Marszewski2024ApJ...967L..41M}. A defining feature of star formation in FIRE is its burstiness, which has been noted as key to how these simulations may be able to reproduce a variety of observational properties \citep{Muratov2015MNRAS.454.2691M, Sparre2016MNRAS.462.2418S, Ma2017MNRAS.466.4780M, Yu2021MNRAS.505..889Y, Stern2021ApJ...911...88S, Gurvich2023MNRAS.519.2598G, Yu2023MNRAS.523.6220Y, Sun2023ApJ...955L..35S}.

In this paper, we study the effect of tidal interactions on the star formation activity of the central galaxy using six zoom-in simulations of Milky Way-mass star-forming galaxies from the FIRE-2 simulation suite \citep{Hopkins2018}. In particular, we analyse the effects of major interactions (stellar mass ratios $\mu$ $\geq$ $0.25$),
minor interactions ($0.1$ $\le$ $\mu$ $<$ $0.25$), and mini interactions ($0.01$ $\le$ $\mu$ $<$ $0.1$) on the star formation histories of the central galaxies. The major interactions in our sample occur at high redshifts, between $z=1.2$ and $z=3.6$. 

The outline of this paper is as follows. In Section \ref{sec:method} we describe the simulations used in the analysis, the halo tracing methods, the SFR calculation and detrending, the classification of galaxy interactions, and the torque calculation; in Section \ref{sec:results} we show the results of this study; in Section \ref{sec:discussion} we discuss the results; in Section \ref{sec:conclusions} we summarize the main findings from this study.

We adopt a standard flat $\Lambda$CDM cosmology with cosmological parameters $H_{0}=70.2 \rm kms^{-1}Mpc^{-1}$, $\Omega _{\Lambda} =0.728$, $\Omega_{m}=1-\Omega_{\Lambda}=0.272$, $\Omega_{b}=0.0455$, $\sigma _{8}=0.807$, and $n=0.961$.

\section{Simulations and Analysis Methods}
\label{sec:method}

In this study, we investigate the causes of starburst events using six zoom-in simulations from the FIRE-2 simulation suite. More specifically, we study what role flybys and mergers play in inducing starbursts. In this section, we describe the simulation sample, the halo identification and tracking method, the SFR and torque calculation, and the detrending of the SFR.

\subsection{The Simulations} 
%We analyze simulations from the FIRE-2 project. With the FIRE-1 simulations using the GIZMO simulation code in ``P--SPH" mode, 
We analyze the FIRE-2 simulations from the FIRE (Feedback in Realistic Environments) project \citep{Hopkins2014, Hopkins2018}, a set of cosmological ``zoom-in'' simulations run with the {\sc gizmo} code \citep{Hopkins2015} \footnote{\url{http://www.tapir.caltech.edu/~phopkins/Site/GIZMO.html}} from redshift 99 to 0. The simulations use a Meshless Finite Mass (MFM) hydrodynamic solver with explicitly modelled star formation and feedback processes. These processes include energy, momentum, mass, and metal fluxes arising from SNe types I\&II, stellar mass-loss (O-star and AGB), radiation pressure (UV and IR), photo-ionization, and photo-electric heating. For more information about the simulations and feedback prescriptions, please refer to \citet{Hopkins2018}.
%The simulations include the cosmic UVB background model of \citet{Faucher2009} together with local radiation sources. Self-shielding is treated with a local Sobolev/Jeans-length approximation. 

In this paper, we study the effect of tidal interactions on the star formation activity around Milky Way-mass star-forming galaxies. To match with the mass of the Milky Way galaxy, we select the galaxies according to their dark matter halo mass at z=0 in the zoom-in simulations from the FIRE-2 simulation suite \citep{Wetzel2023ApJS..265...44W}. The selected galaxies and their characteristics are listed in Table \ref{tab:simulations}. A sub-grid model for turbulent metal diffusion is included for all simulations in our selected sample. The effects of magnetic fields or cosmic rays are not included in the selected runs.

  \begin{table*}
  \begin{threeparttable}
  \centering
   \caption{Simulated galaxies in this study and their parameters.}
	\label{tab:simulations}
	\begin{tabular}{lccccc} 
		\hline
Name & $M_{ \rm halo}^{0}$ & $M_{\rm \star}^{0}$ & $R_{\rm vir}^{0}$ & $m_{\rm dm}$&Reference\\

   &  [$ 10^{12}\,\rm M_{\rm \odot}$]&  [$10^{10}\,\rm M_{\rm \odot}$] & [kpc] &[$ \rm M_{\rm \odot}$]  &        \\
		\hline
$m12w$ & $1.08$ & $5.7$ & 301&  39,000   & 1\\
$m12i$ & $1.18$ & $6.3$ & 311 &  35,000   & 2\\
$m12c$ & $1.35$ & $5.8$ & 328&  35,000 & 3\\
$m12b$ & $1.43$ &$8.5 $ & 331 &  35,000 & 3\\
$m12m$ & $1.58$ &$11$ & 341 &  35,000  & 4 \\
$m12f$ & $1.71$ &$7.9$ & 352 &  35,000 & 5 \\
		\hline
	\end{tabular}
        \begin{tablenotes}
      \small
      \item \textbf{Columns:} 
      \item (1) Name: Simulation designation
      \item (2) $M_{halo}^{0}$: Approximate mass of the $z=0$ ``main'' halo (most massive halo in the high-resolution region).
      \item (3) $M_{\rm \star}^{0}$: Stellar mass of the central galaxy in the main halo at $z=0$. 
      \item (4) $R_{\rm vir}^{0}$: $R_{\rm vir}$ of the $z=0$ ``main" halo as identified through the Rockstar Halo Finder, which corresponds to $R_{360}$ \citep{Behroozi2013}. 
      \item (5) $m_{dm}$: Dark matter particle mass 
      \item (6) References: 1: \cite{Samuel2020}, 2: \cite{Wetzel2016}, 3: \cite{Garrison-Kimmel2019}, 4: \cite{Hopkins2018}, and 5: \cite{Garrison-Kimmel2017}.
      \item We note that all simulations have a initial gas particle mass of 7100 $\rm M_{\rm \odot} $

    \end{tablenotes}
\end{threeparttable}
\end{table*}

\subsection{Identifying and Tracing Halos}
\label{sec:IdentifyingFlybyHalos} % used for referring to this section from elsewhere

To study the effect of galaxy interactions on the star formation activity of the main galaxy, we need to identify and trace the main halo and the companion halos across all simulation snapshots. Each galaxy simulation has 600 snapshots across cosmic time, leading to a snapshot time resolution of 20-25 Myr. We identify the (sub)halos using the ROCKSTAR 6D phase-space temporal halo finder \citep[Robust Overdensity Calculation using K-Space Topologically Adaptive Refinement,][]{Behroozi2013}. ROCKSTAR is based on adaptive hierarchical refinement of friends-of-friends groups in six phase-space dimensions and the time dimension. The halos are identified using dark matter particles, when the average density is 360 times the mean matter density within $R_{360}$ with bound mass fraction $>$0.4, and when there are at least 30 dark matter particles in a halo. Merger trees are constructed using {\it consistent-tree} \citep{Behroozi2013ApJ} with the halo catalog generated by ROCKSTAR for each snapshot. We have compared the output of ROCKSTAR with other halo finders and found similar results, though ROCKSTAR tends to be more robust in tracing subhalos when they are close to the center of a larger halo. Further details on the ROCKSTAR halo catalogs are discussed in \citet{Wetzel2023ApJS..265...44W}. 

Within each zoom-in simulation, the main halo is selected according to its mass at $z=0$ -- the most massive halo in the high-resolution region is identified as the main halo. Among the main halo's progenitors at previous snapshots according to the results from {\it consistent-tree}, the one with the smoothest and generally monotonically increasing mass history is selected as the central halo through all snapshots from $z=3.6$ to $z=0$.

For the potential companion galaxy sample, we select the 1000 most massive halos at $z=3.6$ in each simulation, and trace them to $z=0$, though we note that many fewer halos actually interact with the central galaxy. The large initial selection ensures that all progenitor halos that merge into a galaxy that interacts with the central halo are accounted for. We calculate the stellar mass of the companion galaxies by summing over all star particles within 10 \% of the virial radii as reported by ROCKSTAR.
% As the galaxies merge, there are fewer than 1000 halos at lower redshift. 

\begin{figure}
	\includegraphics[width=\columnwidth,height=\textheight,keepaspectratio]{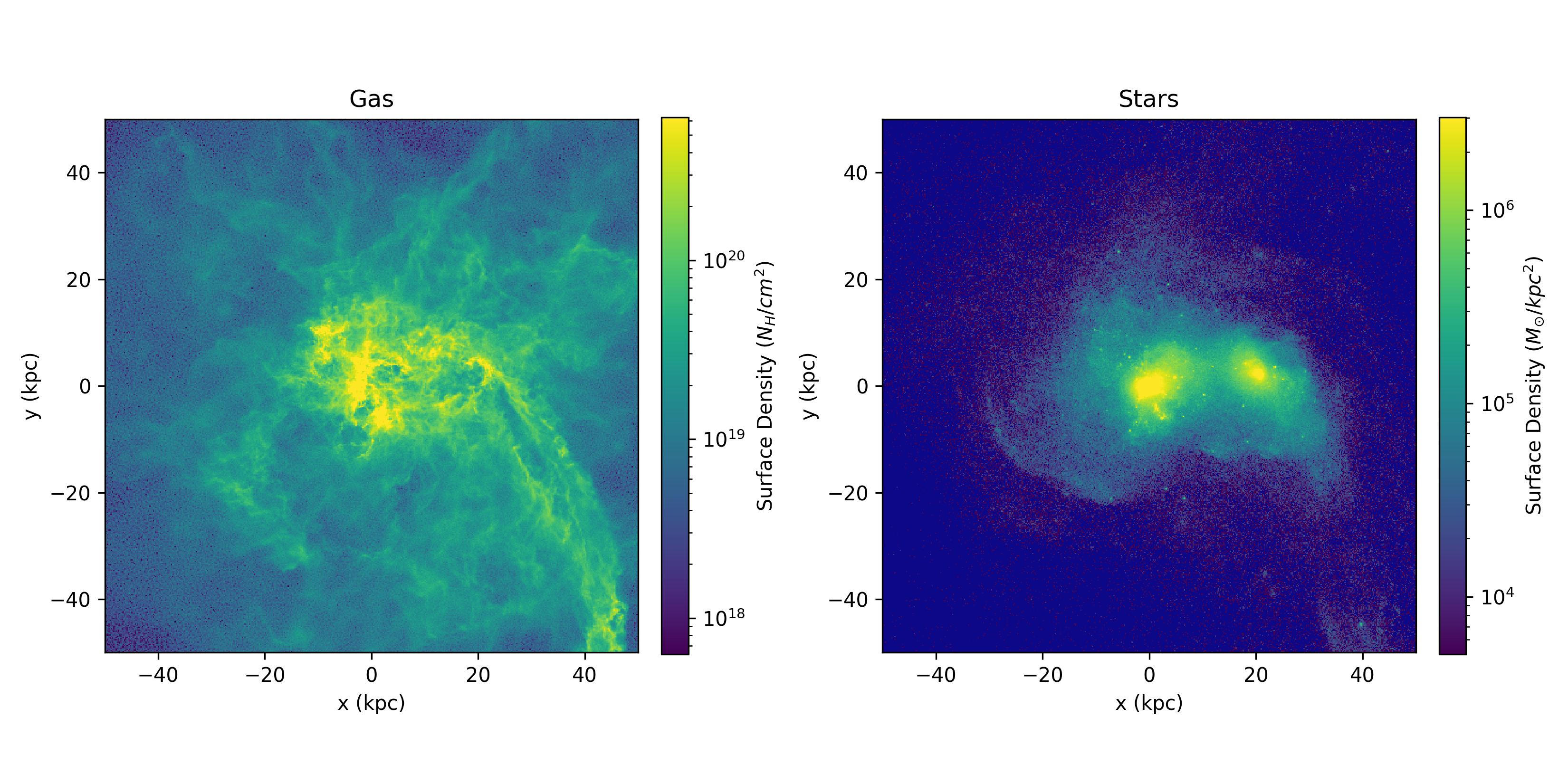}
    \caption{A major merger in the simulation run `m12m' at $z=1.6$ with a stellar mass ratio of 0.66. The distance between the two galaxies is 22 kpc. The left panel shows the gas distribution of the central and companion galaxies, and the right panel shows the stellar component of the galaxies.}
    \label{fig:frame_m12b}
\end{figure}

\begin{figure}
% figure from Tracing/plotting/Distance_redshift.ipynb
	\includegraphics[width=\columnwidth,height=\textheight,keepaspectratio]{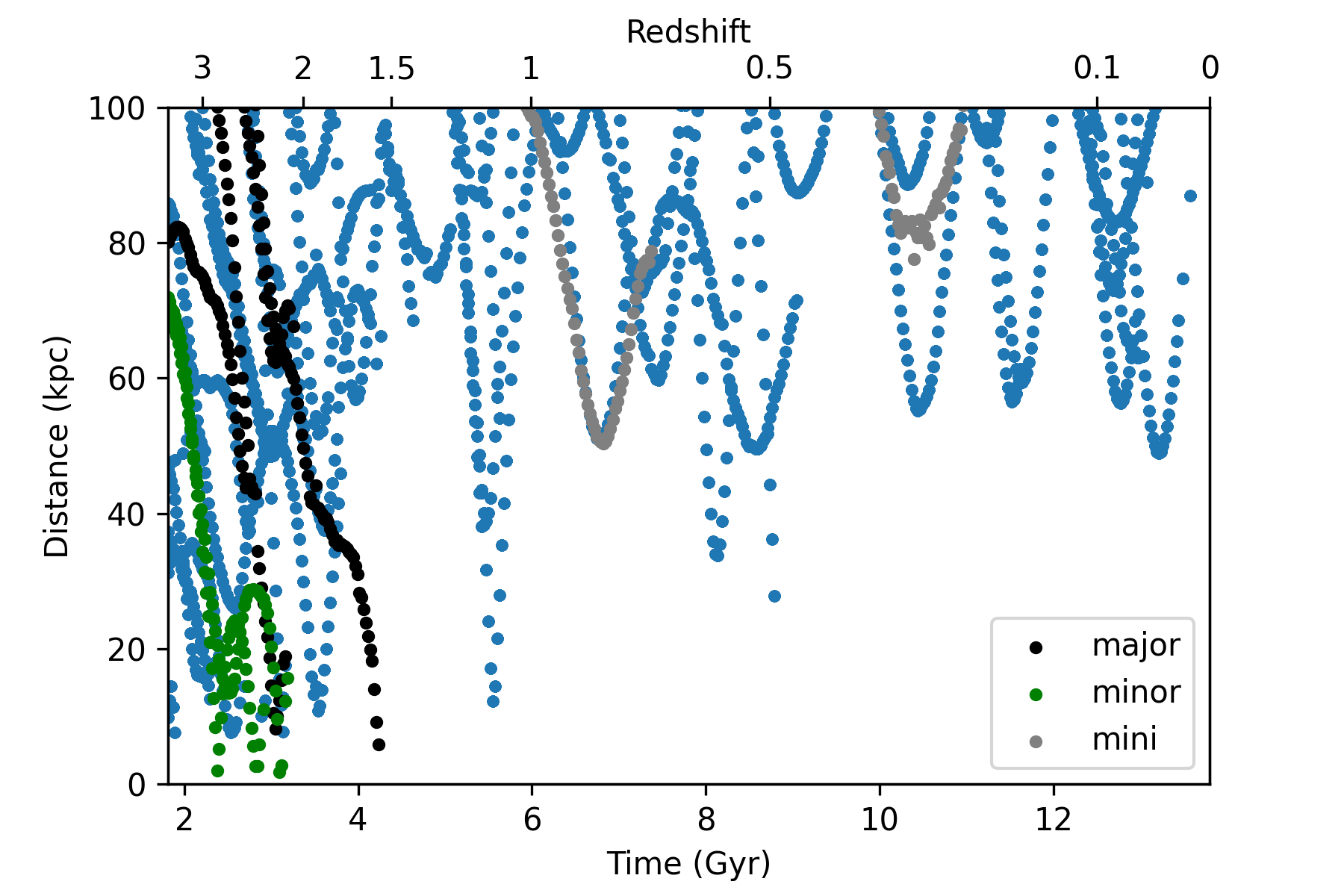}
    \caption{The distance of all the nearby halos with dark matter mass above $10^{9} M_{\odot}$ to the center of the main halo as a function of cosmic time for the simulation run `m12m'. The black dots show the distance of galaxies in snapshots with major interactions, where the companion-to-central galaxy stellar mass ratio is above or equal to 0.25; the green and grey dots represent galaxies involved in minor ($0.1$ $\le$ $\mu$ $<$ $0.25$) and mini ($0.01$ $\le$ $\mu$ $<$ $0.1$) interactions, respectively; while the blue dots show galaxies with stellar mass ratios smaller than 0.01 (but halos masses larger than $10^9\, M_\odot$). The $R_{\rm vir}$ is 107 kpc at $z=3.6$, and 341 kpc at $z=0$.}    
    \label{fig:distance_m12m}
\end{figure}

Figure \ref{fig:frame_m12b} shows an example of the distribution of the gaseous and stellar component of the main halo in the simulation run `m12m' at $z=1.6$, where the central galaxy is undergoing a major merger. We present an example of the companion galaxy identification results in Figure \ref{fig:distance_m12m}: we plot the distance of all the nearby halos with dark matter mass above $10^{9} M_{\odot}$ from the center of the main halo as a function of cosmic time for the simulation run `m12m'.

\subsection{Identifying and Categorizing Halo Interactions} \label{sec:categorizing_interactions}

With the primary halo identified and traced across all simulation snapshots, we are able to filter our investigation to only halos that interact with the main galaxy for further analysis. These halos are identified by finding galaxies with a stellar mass greater than 1\% of the simulations' primary galaxy that approach within 100 kpc of it. The population of satellites selected using this criteria is broadly similar to the satellites selected if total mass was used as the criteria. We note that the tracing of halos close ($\sim$ kpc) to the central galaxy is challenging given the confusion of particles in the snapshot, so we limit the tracing of these subhalos to 10 kpc from the center of the main halo.  

The identified galaxies may either merge into the main galaxy after their final approach, or simply flyby the main galaxy with minimal exchange of mass. For the purpose of this study, we deem both of these cases to be interactions. We do note, however, that all galaxies in these simulations with a stellar mass ratio of $\mu \ge 0.25$ that approach the main galaxy closer than 100 kpc, do eventually merge within 1-2 Gyr.

We further subdivide the interactions into three mass ratio-related categories \footnote{In this paper, we use the mass ratio calculated at each snapshot. There are other studies of galaxy simulations \citep[e.g.,][]{Cox2008, Martin2017} that use the maximum mass of the secondary galaxy prior to coalescence for the calculation of the mass ratio for all snapshots in which the merging companions are involved in.}, the \textbf{major interactions}, with stellar mass ratios of $\mu \ge 0.25$, the \textbf{minor interactions}, with $0.1 \le \mu < 0.25$, and the \textbf{mini interactions}, with $0.01 \le \mu < 0.1$. We use these classifications to enable fine-grained analysis of the effect of the interactions.

Each of the snapshots across all simulations are classified based on if the central galaxy is undergoing an interaction by the criteria above. We note that more than one galaxy could meet the criteria for a given snapshot, but the mass-ratio classification is based on the most massive of those galaxies. These classifications are further verified manually to remove any misclassifications due to confusion in companion halo parameter measurement: for instance, removing artificially inflated masses of close-passing satellites where simulation particles from the main galaxy are erroneously associated with the satellite. Across the six simulations, we identify that the main galaxy is undergoing some form of interaction $\sim$40\% of the time.

% The snapshots in the simulations are divided into four groups according to the ratio of the stellar masses of the satellite galaxy and the central galaxy. To determine which merger group a certain snapshot belongs to at each snapshot, we use the satellite galaxy with the highest stellar mass within 100 kpc of the centre of the central galaxy, and take the ratio of the stellar mass of the satellite galaxy to the stellar mass of the central galaxy. The four merger types are major mergers (stellar mass ratios $\mu$ $\geq$ $0.25$),
% minor interactions ($0.1$ $\le$ $\mu$ $<$ $0.25$), mini interactions ($0.01$ $\le$ $\mu$ $<$ $0.1$), and others ($\mu$ $<$ $0.01$ or without satellite galaxies within 100 kpc). There are cases of satellite galaxies artificially increasing in stellar mass when very close to the central galaxy as stars from the central galaxy are mistakenly attributed to the satellite; we have manually checked all mergers to remove the falsely identified major mergers.

It is worth noting that the zoom-in simulations used in this study were selected only to match the halo mass of the Milky Way at $z=0$ and not specifically selected to match any other property of the Galaxy, including its relative quiescence at low redshift \citep{Wetzel2023ApJS..265...44W}. Despite that, all major mergers in the sample typically occur at early times ($z >1.2$), consistent with the Milky Way's decreasing merger rate at late times \citep[e.g.,][]{Conselice2003AJ....126.1183C, Bell2006ApJ...652..270B, Lackner2014AJ....148..137L}. Further, we also note that the minimum distance of the satellite galaxy approaches increases at late times. We discuss these properties in more detail in \S \ref{sec:results}. 

\subsection{Measuring Star Formation}
\label{sec:detrending}

To study the effect of the tidal interactions on the central galaxy, we need to know the exact times that starbursts occur in the galaxies. We calculate the short-term star formation rate of each snapshot by summing over the mass of stars formed within 5 Myr, corresponding to the age at which the massive stars producing ionizing radiation evolve off the main sequence, and divide the total mass of newly formed stars by the time interval (5 Myr). We do so for all stars within 10 kpc of the center of the galaxy, consistent with the size of the stellar disk. 

As a check of the robustness of our calculated SFR, we have performed the same calculation with radii up to 50 kpc, and for stars formed within different time intervals ($<20$ Myr, the default time between snapshots), and found that the SFR history is generally not sensitive to the choice of the radius or the time interval.

The raw SFR has large-scale secular variations on several hundred million to billion year time scales. To determine the short time ($\sim$10 Myr) effect of interactions, which happen on time scales of $10-100$ Myr \citep[e.g.,][]{Keel1985, Barnes1991, Hopkins2010MNRAS.402..985H, Cenci2024MNRAS.527.7871C}, we also utilize a detrended SFR.
We detrend the SFR by calculating a smoothed SFR with a Gaussian filter with $\sigma = 25$ snapshots ($\sim$ $550$ Myr), then divide the original raw SFR by the smoothed to provide a unitless over/under abundance of star formation as compared to the secular trend. This makes it possible to compare the results across different redshifts and between different galaxies.   

\subsection{Measuring Interaction Torques}
\label{sec:method_torque}
%we perform the torque calculation using all the satellite galaxies we trace. 

It has been suggested that one possible mechanism for the enhancement of star formation from galaxy interactions is the torque exerted by the stars in the central galaxy on the gas in the same galaxy, as a result of the perturbation by the companion galaxy \citep{Mihos1994, Mihos1996, Barnes1996, Hopkins2013}. The mutual torque between the stars and gas in the main galaxy leads to a decrease in the angular momentum of the gas, and thus induces gas inflow to small radii, thereby enhancing the SFR. 

The calculation of the torque between the stellar and gaseous components for all our snapshots for all our central galaxies is computationally prohibitively expensive. In this work, we used the torque from the companion galaxies on the gas particles in the central galaxy as a proxy; we assume this torque is proportional to or has a monotonic relationship with the torque from stars to the gas in the central galaxy.

To quantify the impact of interactions on the star formation activity in the main galaxy, we calculated the torque exerted on the gas particles in the central halo from the nearby halos. The galaxies within $1500$ kpc of the main halo at each snapshot are included in the calculation. The expression of the torque is $\vec{r} \times \vec{F}$, where $\vec{r}$ is the vector from the center of the main galaxy to the gas particle in the disk, and $\vec{F}$ is the gravitational force vector, 
\begin{equation}
{\vec F} = \frac{GM_{\rm f}m_{\rm i}}{R^{3}}\vec{R} ,
\end{equation} 
where $G$ is the gravitational constant, $M_{\rm f}$ is the dark matter mass of the flyby halo as reported by ROCKSTAR, $m_{\rm i}$ is the mass of the gas particle in the main galaxy, and $\vec R$ is the vector from the gas particle to the center of the flyby galaxy. We perform the calculation by summing over the torque from the companion halos on every gas particle in the main halo,
%\begin{equation}
%\vec{\tau} = \vec{r} \times \vec{F}
%\end{equation} 
\begin{equation}
\vec{\tau} = \sum_{M_{\rm f}} \sum_{m_{\rm i}} \vec{r} \times \frac{GM_{\rm f}m_{\rm i}}{R^{3}}\vec{R}. 
\end{equation} 
We calculated this torque in spherical shells of width $\Delta R=1\,\kpc$ from the center of the main galaxy to $R=20\,\kpc$. At large radii, this procedure will include gas not in the galactic disk. We did this for each snapshot for all the galaxies we discuss. We also use specific torque in our analysis, the specific torque is calculated as the torque divided by the total gas mass in the region.

We note that the torque calculated at each snapshot through this procedure is the sum of the torque of all halos within 1.5 Mpc on the central galaxy. However, these torques are dominated by the effect of one or two halos for any given snapshot.

\section{Results}
\label{sec:results}

%\subsection{Torque and The Detrended SFR}
%\label{sec:torque_and_SFR}

% Overall Picture of M12 Galaxies over cosmic time
% Note about how mergers sometimes bring in gas, but don't seem to affect the overall star formation rate

\begin{figure}
\includegraphics[width=\columnwidth,height=\textheight,keepaspectratio]{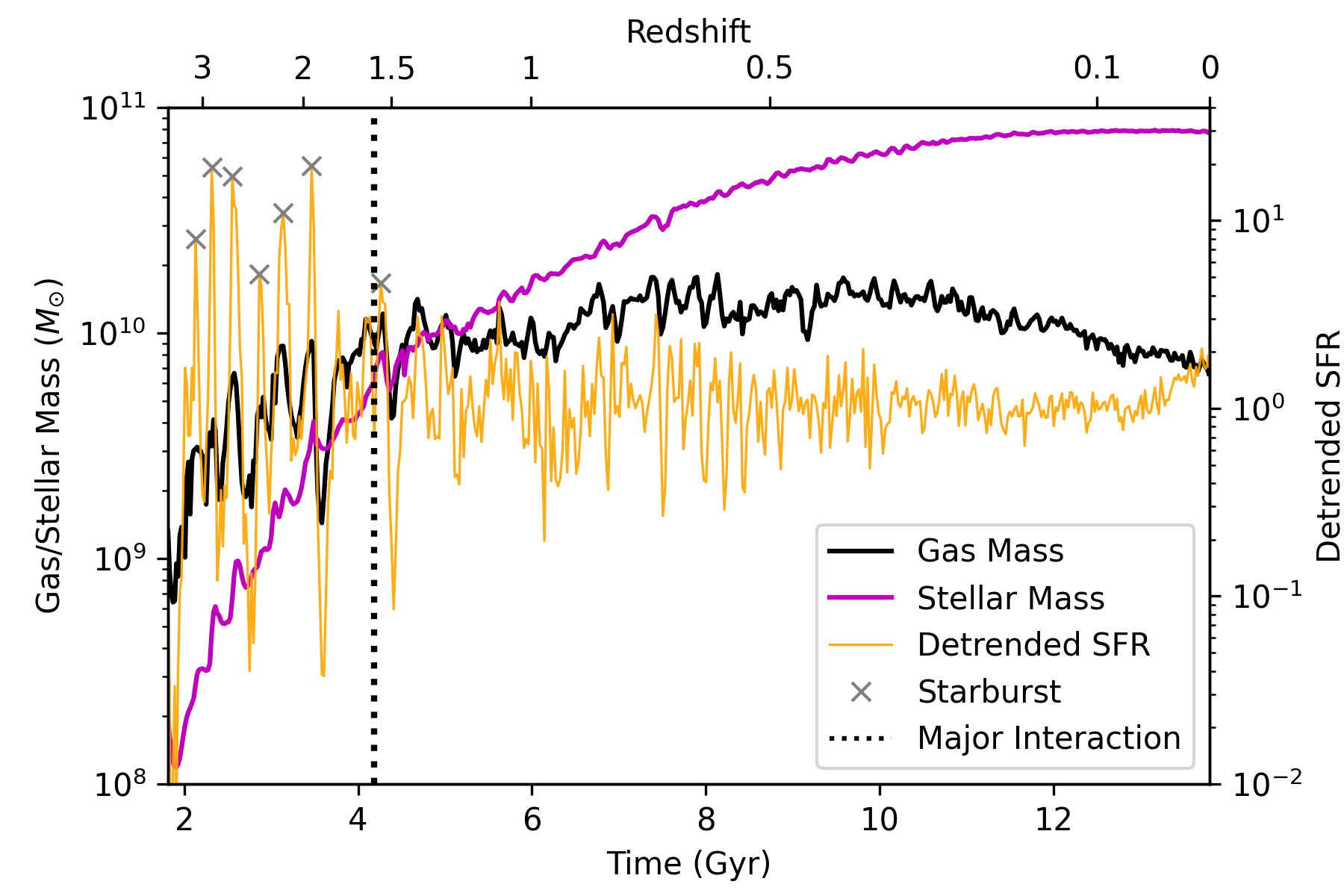}
    \caption{The gas mass (the black line), the stellar mass (the magenta line), and the detrended SFR (the orange line) of the central galaxy in `m12m' as a function of cosmic time. The black dotted vertical line marks the end of a major merger event, which is defined as when the centre of the satellite galaxy is within 10 kpc of the central galaxy. The gray x's mark starburst events, where the detrended star formation rate is greater than 4. There are seven such starburst events; during the same period, there is one major merger and 3 minor interactions, suggesting that most of the starbursts are not associated with major or minor interactions. All of the simulated galaxies in the sample have mass and star formation histories that are qualitatively similar with those shown here. Similar plots for other sample galaxies are shown in Figure \ref{fig:gas_star_dSFR_m12c}, and in Figures \ref{fig:gas_star_dSFR_m12w} and \ref{fig:gas_star_dSFR_m12f} in Appendix \ref{app_c}.}
    \label{fig:gas_star_dSFR_m12m}
\end{figure}

\begin{figure}
\includegraphics[width=\columnwidth,height=\textheight,keepaspectratio]{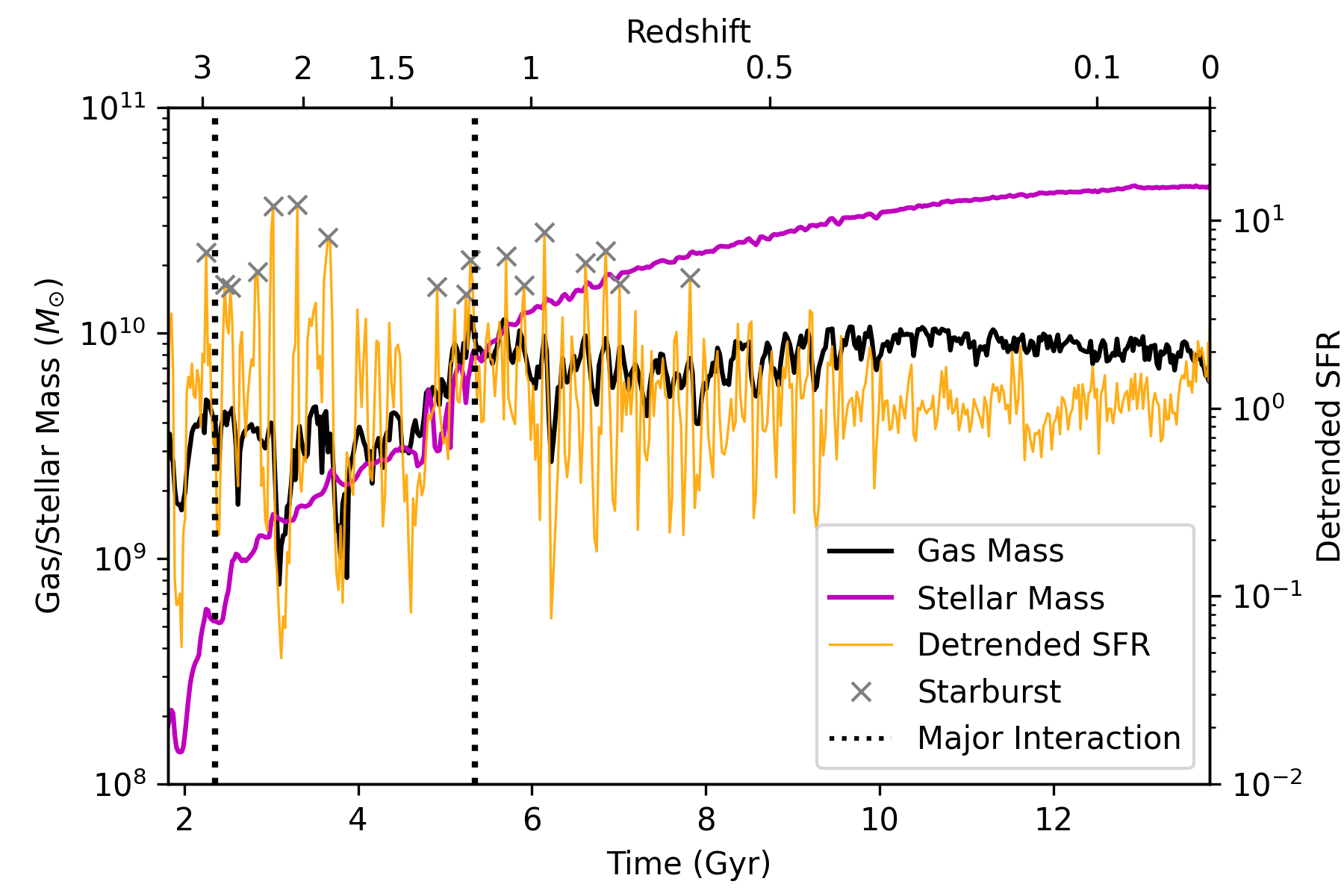}
    \caption{Similar to Figure \ref{fig:gas_star_dSFR_m12m}, but for the simulated galaxy `m12c'. This is the only galaxy in the sample with a measurable increase of the gas mass due to a major merger. We note that the detrended star formation history appears similar to all other galaxies within our sample. We see 17 starbursts over the redshift range of the figure, while there are 2 major mergers and 2 minor interactions, suggesting that many of the starbursts are not caused by significant interactions.}
    \label{fig:gas_star_dSFR_m12c}
\end{figure}

The FIRE m12 series of galaxy simulations provide a consistent picture of the interaction and star formation history of Milky Way-mass galaxies. While the only selection criterion applied to these galaxies is that they must match the halo mass of the Milky Way at $z = 0$, they all appear to have similar behavior (Figures \ref{fig:gas_star_dSFR_m12m} \& \ref{fig:gas_star_dSFR_m12c}): All major mergers occur at early times, between $1.2 < z < 3.6$. The merger rate of the galaxies decreases with redshift at late times \citep[e.g.,][]{Hopkins2010ApJ...724..915H, Mantha2018MNRAS.475.1549M, Rodriguez2015MNRAS.449...49R}. The star formation in the galaxies goes from bursty to more stable at $z \sim 1$, which also coincides with the onset of disk formation, and the disks then become thinner over time, e.g., \citet{Yu2021MNRAS.505..889Y, Hafen2022MNRAS.514.5056H, Yu2023MNRAS.523.6220Y,  McCluskey2024MNRAS.527.6926M}. For the mergers identified through these simulations, some are able to bring gas into the main halo (i.e. `gas-rich' mergers), while others do not (i.e. `gas-poor' mergers), but in either of these cases, most mergers do not have a significant effect on the overall star formation rate of the galaxy, nor does it consistently lead to a burst in star formation at the time of the merger. For Milky Way-mass halos, it appears that the time, size, or gas-richness of the mergers have no significant impact on the overall star formation history or present day star formation rate of the galaxy. We also note that the duration of the starbursts (when the detrended SFR is greater than 4) are 25-50 Myrs.

% The zoom-in simulations in this study were selected to match the halo mass of the Milky Way at $z=0$. The major mergers in the sample typically happen at early times, between $z=1.2$ and $z=3.6$, with the latest one happening between $z=1.9$ and $z=1.2$ for simulation `m12c'. This is consistent with the fact that the merger rate decreases with redshift at late times. 

% Correlations between Star Formation and Torques
% Maybe start with something about how there's not an obvious visual correlation before going into correlation coefficients

\subsection{Do Galaxy Interactions Trigger Starbursts?}

While we do not see a significant impact on the overall star-formation history due to interactions, they may yet be responsible for catalyzing specific starburst episodes within the host galaxy. Specifically, the torque on the system due to the companion galaxy may produce a local starburst (or otherwise elevated star formation rate) at the time of closest approach. We approach this proposed starburst mechanism by asking the following question: Does the star formation rate of a galaxy increase with the torque exerted on it from a companion galaxy?

To address this question, we measure the correlation between the detrended SFR of the main halo in each of these simulations and the torque exerted on the halo by any companion galaxy. We adopt a detrended SFR to minimize any statistical contamination from secular trends in the star formation history of the galaxy, as described in Section \ref{sec:detrending}. We adopt the Spearman rank correlation coefficient as our metric of correlation, which provides a measure of the strength of correlation of two variables based on their rank within the sample. The calculated p-value from the Spearman correlation coefficient indicates the probability of the null hypothesis, that is, the measured correlation occurring by chance. The Spearman correlation test has the benefit of being independent of the scaling of the two variables, being only a function of the rank order of the measurement within the samples. This is in contrast to other statistical correlation tests, such as the Pearson correlation test, which measures the linear correlation between the two variables. We have repeated our analysis with a Pearson correlation coefficient and find no significant differences in the overall results. We note that these correlation tests are designed to measure monotonic relationships, so more complex relationships between the variables would not be uncovered through this study.

\begin{figure*}
% figure from /GIZMOUtils.jl/Tracing/plotting/Specific_Torque_SFR_all_halos.ipynb
	\includegraphics[width=\textwidth]{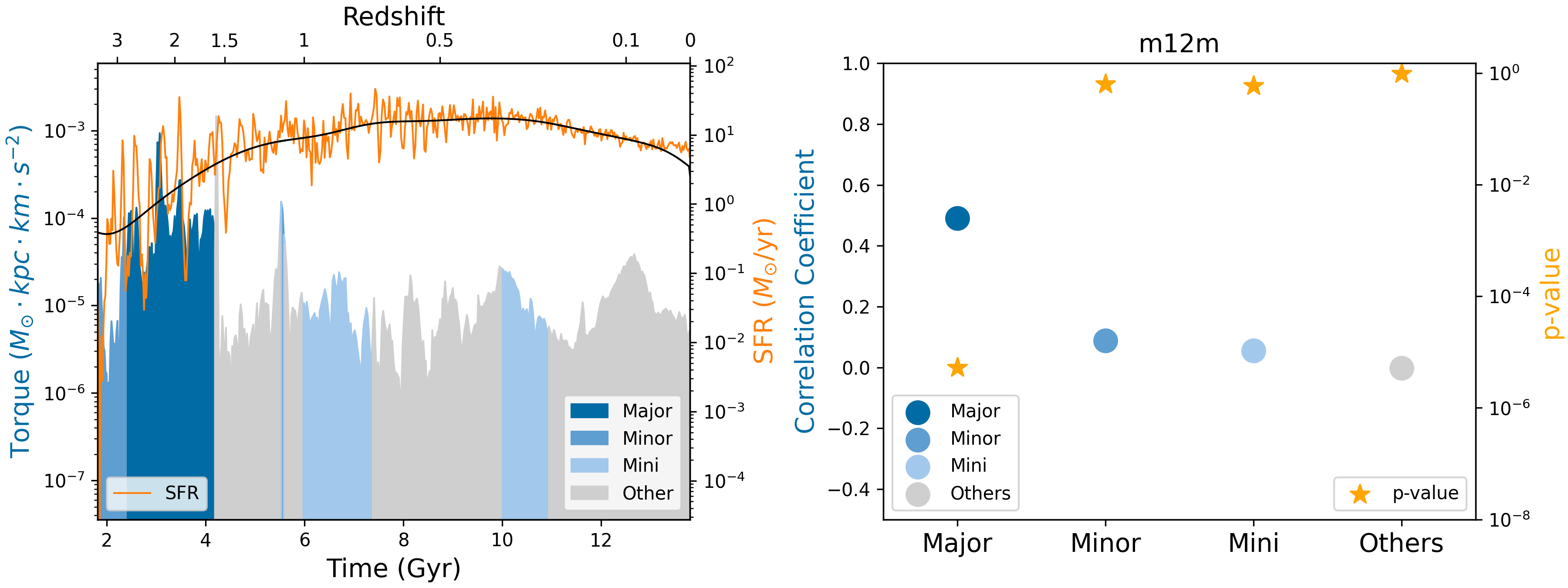}
    \caption{{\em Left:} The torque and SFR for the simulated galaxy `m12m'. The torque is shown as shaded regions, assigned colours based on if they are undergoing an interaction during that snapshot. Dark blue indicates snapshots with major interactions ($\mu$ $\geq$ $0.25$), medium blue with minor interactions ($0.1$ $\le$ $\mu$ $<$ $0.25$), and light blue with mini interactions ($0.01$ $\le$ $\mu$ $<$ $0.1$). Light grey regions indicate no interaction. The SFR is plotted with the orange line, and the calculated smoothed SFR is plotted in black. \\
    % The torque is exerted by satellite galaxies on gas particles within 10 kpc of the central galaxy; and the SFR is calculated for stars less than five million years (stellar mass divided by 5 Myr) within 10 kpc of the centre of the galaxy. The SFR transitions from a `bursty' star formation state to a `steady' star formation state. 
    {\em Right:} the Spearman correlation coefficient between the torque and the detrended SFR for snapshots within major interactions, minor interactions, mini interactions, and other snapshots, shown as the blue and gray dots; the orange stars represent their corresponding p-values.}
    \label{fig:SFR_specific_torque_m12m}
\end{figure*}

On the left of Figure \ref{fig:SFR_specific_torque_m12m}, we present the sum of the torque exerted by the 50 nearby halos that exert the greatest gravitational force on the central galaxy for each simulation snapshot between $0 < z < 3.6$ for `m12m'. We note that although we calculate the torque for the top 50 halos, it is dominated by one or two halos for any given snapshot, due to the $R^{-2}$ relationship between the companion galaxy distance and the torque. We also show the true and smoothed SFR on the same axis for comparison. Similar figures are available for all remaining simulations in Appendix \ref{app_a}.

\begin{deluxetable}{cccccccc} %Table 2
\tabletypesize{\scriptsize}
% \centering
   \tablecaption{The number of major interactions, minor interactions, and starbursts in each simulation; the last two columns show the fraction of time when the galaxies are in starburst mode. Significant interactions cannot account for all or even most of the starbursts. There is no starburst after $z=0.4$, and on average galaxies are in starburst mode for $3.1\%$ of the time in $0<z<3.6$.
    \label{tab:starburst_number}
}

\tablehead{
    Simulation &
    $M_{ \rm \star}^{0}$ &
    Major &
    Minor &
    Starbursts & 
    \multicolumn{2}{c}{Starburst Fraction} &
    Last Major \\
    &
    [$ \rm M_{\rm \odot}$]& 
    Interactions & 
    Interactions & 
    &
    ($0.4<z<3.6$) &
    ($0<z<3.6$) &
    Interaction Redshift
}

 \startdata
m12w & $5.7 \times 10^{10}$ & 0 & 6 & 7 & 2.6\% & 1.6\% & \nodata \\
m12i & $6.3 \times 10^{10}$ & 2 & 3 & 11 & 4.8\% & 3.0\% & 2.0\\
m12c & $5.8 \times 10^{10}$ & 2 & 2 & 17 & 7.4\% & 4.7\% & 1.2\\
m12b & $8.5 \times 10^{10}$ & 1 & 3 & 13 & 5.8\% & 3.7\% & 2.4\\
m12m & $1.1 \times 10^{11}$ & 1 & 3 & 7 & 5.8\% & 3.7\% & 1.6\\
m12f & $7.9 \times 10^{10}$ & 0 & 0 & 9 & 3.2\% & 2.0\% & \nodata \\
 \enddata
\end{deluxetable}

Visually, the number of spikes in the torque exerted by companion galaxies is fewer than the spikes in star formation rate, immediately indicating that the interaction with companion galaxies cannot be solely responsible for the starbursts within a central galaxy. In the literature starbursts are often defined as having 2-4 times the SFR above the galaxy main sequence \citep[e.g.,][]{Rodighiero2011, Schreiber2015, Ellison2020}, $SFR_{starburst}>4\times SFR_{GMS}$. Another selection criterion used is the birthrate parameter \citep{Kennicutt1983}, $b=SFR/\langle SFR \rangle$, i.e. the ratio between the current SFR and the mean SFR over the lifetime of the galaxy, and a galaxy is a starburst galaxy when $b \ge 3$ \citep{Bergvall2016}. In Figures \ref{fig:gas_star_dSFR_m12m}, \ref{fig:gas_star_dSFR_m12c}, and appendix Figures \ref{fig:gas_star_dSFR_m12w} and \ref{fig:gas_star_dSFR_m12f}, we mark the peaks in detrended SFR when the star formation rate is 4 times, or more, than the smoothed SFR. There are on average 11 starburst events per galaxy from $z=3.6$ to $z=0$ in our sample. On average, there are one major interaction and three minor interactions for each galaxy; the number of major/minor interactions and starburst events for each galaxy is shown in Table \ref{tab:starburst_number}. Significant interactions cannot account for all or even most of the starbursts in the star formation history of the FIRE galaxies we examine.

\begin{deluxetable}{LCCCCC} % Table 3
\tablehead{\colhead{Simulation} & \colhead{Major Merger 1} & \colhead{Major Merger 2}
& \colhead{Minor Interactions} & \colhead{Mini Interactions} & \colhead{Others} }
\tablecaption{Spearman correlation coefficients between the torque and the detrended SFR for snapshots involving major interactions, minor interactions, and mini interactions for the 6 simulations, with the p-values for the correlation coefficients shown in the parentheses.\label{tab:correlation_torque_spearman_detrended}}
\startdata
m12w & NA & NA & -0.21 ( $4\times 10^{-3}$ ) & 0.26 ( $2\times 10^{-3}$ ) & 0.15 ( $4\times 10^{-2}$ ) \\
m12i & 0.76 ( $2\times 10^{-7}$ ) & 0.85 ( $10^{-6}$ ) & 0.12 ( 0.46 ) & 0.27 ( $3\times 10^{-3}$ ) & -0.12 ( $5\times 10^{-2}$ ) \\
m12c & 0.55 ( $3\times 10^{-3}$ ) & $-7\times 10^{-2}$ ( 0.58 ) & 0.00 ( 1.00 ) & 0.11 ( 0.39 ) & $1\times 10^{-1}$ ( $8\times 10^{-2}$ ) \\
m12b & $9\times 10^{-2}$ (0.57) & NA & -0.14 (0.19) & 0.22 (0.31) & $-9\times 10^{-2}$ ( 0.10 ) \\
m12m & 0.50 ( $3\times 10^{-6}$ ) & NA & 0.12 ( 0.52 ) & $9\times 10^{-2}$ ( 0.36 ) & $5\times 10^{-2}$ ( 0.36 ) \\
m12f & NA & NA & NA & 0.19 ( $6\times 10^{-3}$ ) & $-5\times 10^{-2}$ ( 0.38 ) \\
\enddata
\end{deluxetable}

We present the Spearman correlation coefficients (and related p-values) for the relationship between torque and detrended star formation for `m12m' on the right of Figure \ref{fig:SFR_specific_torque_m12m}. We note that the results of this correlation test are similar if we use the true SFR rather than the detrended, or if we use a Pearson correlation test rather than a Spearman test. Within the `m12m' simulation, we find there is no significant correlation between torque and SFR for snapshots, other than for major mergers, where a moderate but significant correlation does exist. This is similar for other galaxies, where there is a correlation of the SFR with four out of six major mergers. There are no major interactions in `m12w' and no major or minor interactions in `m12f' over the time interval we examined. Over the same time, the number of starbursts in `m12w' is five, and in `m12f', nine. The same plot for the other simulated galaxies is shown in Figures \ref{fig:SFR_specific_torque_m12b} and \ref{fig:SFR_specific_torque_m12w} in Appendix \ref{app_a}.

The correlation coefficients and their p-values for each type of interaction and each simulation are shown in Table \ref{tab:correlation_torque_spearman_detrended}. For major interactions, the detrended SFR has high correlation coefficients ($\ge 0.5 $) for 4 of the 6 simulated galaxies; for minor and mini interactions, there is a statistically significant correlation coefficient for 4 of the 12 cases, and the correlation coefficient is around 0.2 for those cases; and for other snapshots, there is no information on the correlation between the SFR and the torque, as the p values are all high. This indicates that in some cases, major interactions are correlated with starbursts, but starbursts are more frequent than major interactions, and occur even in the absence of major or minor interactions.

The snapshots across the different simulations can be stacked by having the torque and detrended SFR values normalized in logarithmic space using the maximum and minimum values within each interaction type in each simulation. We can then perform the analysis using all snapshots with the different types of interaction in all simulations. Table \ref{tab:correlation_combined} shows the correlation coefficient from the stacked snapshots. For major interactions, there is a positive and statistically significant correlation between the torque and the detrended SFR for major interactions, with a Spearman correlation coefficient of 0.41, and a p-value of $2.0 \times 10^{-12}$; for minor interactions, mini interactions, and other snapshots, the p-values for the correlation coefficients are large and therefore the correlation coefficients are not statistically significant.

\begin{table} % Table 4
%  \begin{threeparttable}
%	\centering
   \caption{Spearman correlation coefficients between the torque and the detrended SFR for stacked snapshots involving major interactions, minor interactions, mini interactions, and other snapshots for the 6 simulations, with the p-values for the correlation coefficients shown in the third column. The torque and the detrended SFR are normalized in logarithmic space according to the maximum and minimum values within each interaction group in each simulation run.}
	\begin{tabular}{lcr} 
		\hline
Interaction Type & Correlation Coefficient & p-value \\
 \hline
 
Major Interactions & 0.41 & $2.0 \times 10^{-12}$ \\
Minor Interactions & $0.11$ & 0.04 \\
Mini Interactions & 0.10 & 0.01 \\
Others & 0.05 & $0.05$ \\

 \hline
 \label{tab:correlation_combined}
 %\caption{Pearson correlation coefficient between the torque and the SFR for different snapshot bins and spherical shells for the simulation run `m12w'. The p-values for the correlation coefficients are shown in the parentheses.}
	\end{tabular}
   %     \begin{tablenotes}
   %   \small
   %   \item     

    %\end{tablenotes}
%\end{threeparttable}
\end{table}

\begin{figure*}
% figure from /GIZMOUtils.jl/Tracing/StarMassRatio/combined_major.ipynb
	\includegraphics[width=\textwidth]{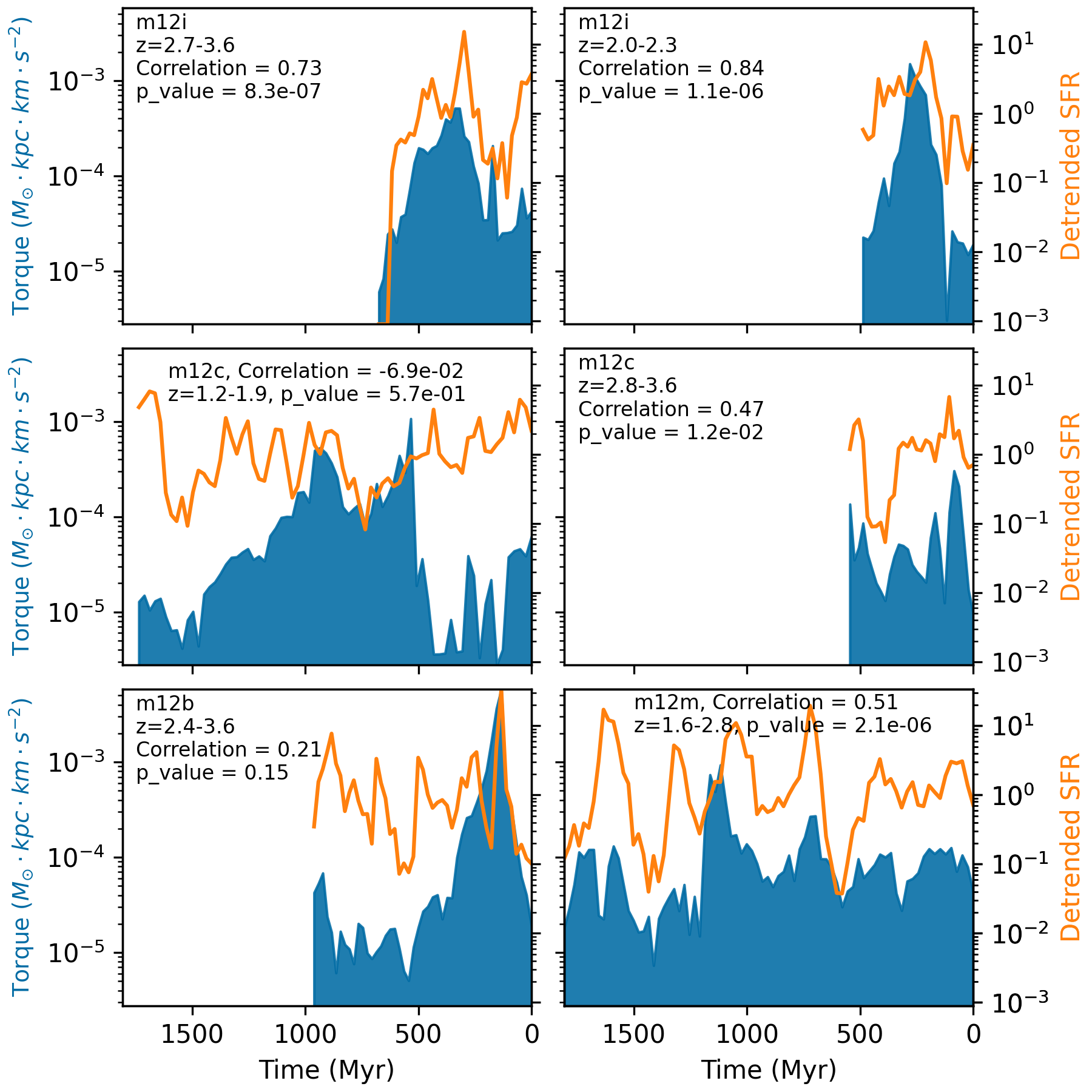}
    \caption{The torque and the detrended SFR during major interaction snapshots in the 6 simulation runs. The x axis shows the time until the final merger. In the analysis a merger starts when a halo with stellar mass ratio $\mu$ $\geq$ $0.25$ is within 100 kpc of the center of the central galaxy, and ends when the companion galaxy is within 10 kpc of the center of the central galaxy and does not re-appear outside that radius as a separate halo again. Each major merger is shown as a single panel, some panels involve multiple major mergers occurring close in time. There are 6 major mergers in all the simulated galaxies. The simulated galaxies `m12f' and `m12w' do not have any major interaction between $z=3.6$ and $z=0$.}
    \label{fig:major_mergers}
\end{figure*}

Figure \ref{fig:major_mergers} shows the torques and star formation rates of all the major interactions in all the simulations we analyzed. Three of the major interactions have a high correlation ($>0.5$) between torque and detrended SFR; while other major interactions show a low or no correlation between the two quantities. In comparison, the correlation between the gas mass increase rate and the detrended SFR is around 0.1 for all simulations. Consistent with our finding above, it appears that major interactions cause the starbursts in some cases, but are not responsible for all or even most of the starbursts. This is expected as each burst will cause subsequent gas infall and potentially more bursts \citep{Muratov2015MNRAS.454.2691M, Angles2017MNRAS.470.4698A}.

\begin{figure}
% figure from 
\centering
\includegraphics[width=\columnwidth,height=\textheight,keepaspectratio]{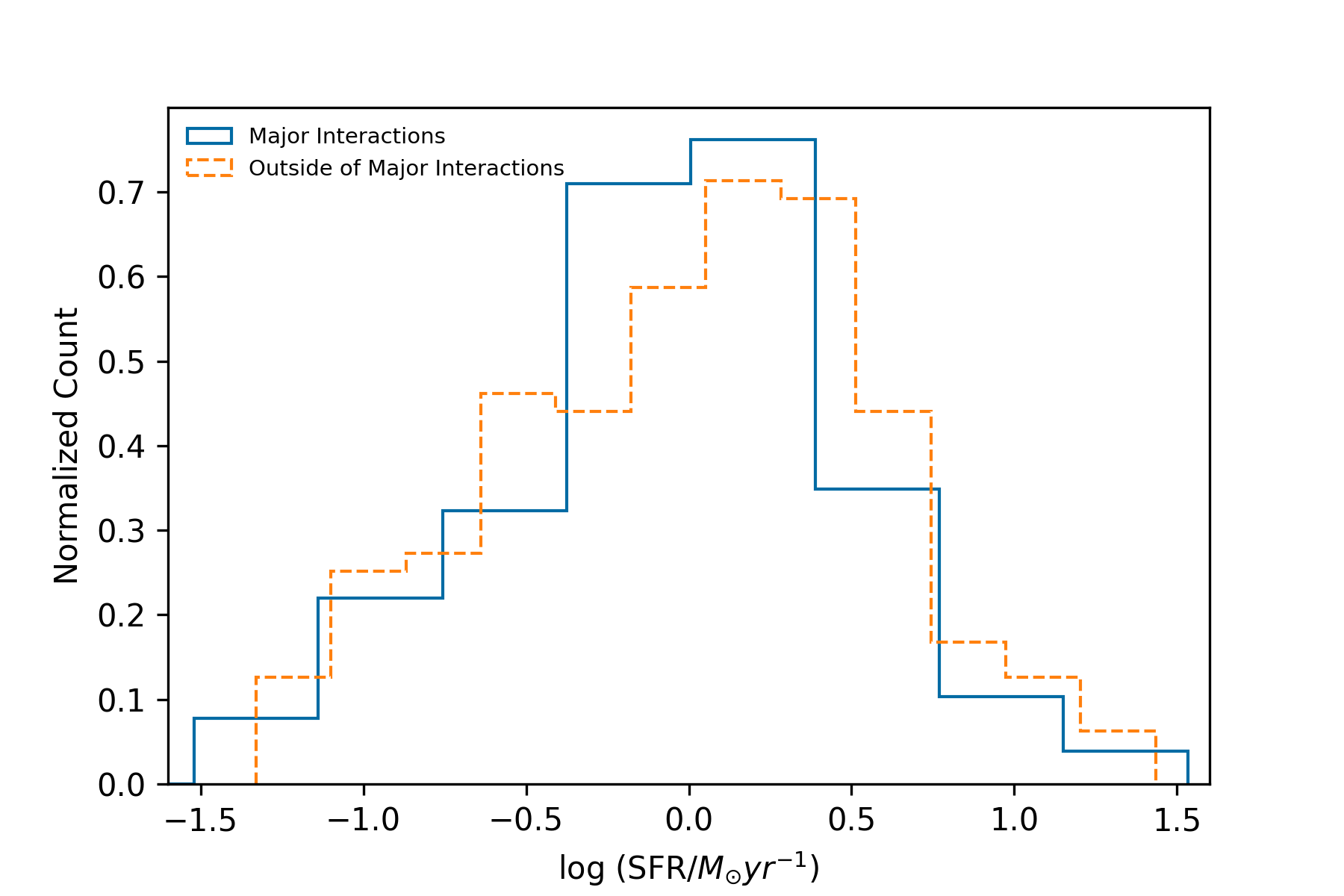}
    \caption{The normalized histogram of the detrended SFR for snapshots during a major interaction (blue, solid), and 0.5 Gyr before and after the interactions (orange, dashed). The two distributions show no significant difference, indicating that there is no statistical enhancement of the star formation rate at the time of major interactions as compared to outside of these major interactions.
    }
    \label{fig:histogram_SFR}
\end{figure}

In Figure \ref{fig:histogram_SFR}, we present the distributions of the detrended SFR from snapshots during major interactions, using snapshots up to 0.5 Gyr before and after the interaction as a control sample. We note that for the control sample, snapshots within the 0.5 Gyr time window where another major interaction is occurring have been excluded. If there was an enhancement in the star formation rate during major interactions, we would expect to see the distribution of the detrended SFR to be shifted for the major interaction sample. We find no significant difference between the two samples using a K-S test (p-value of 0.14), indicating that we find no measurable enhancement of the star formation during major interactions. We note that this analysis is insensitive to the distance criteria used for when an interaction begins (i.e., 100 kpc as discussed in section \ref{sec:categorizing_interactions}), with the distributions being similar when the distance criteria is limited to 50 kpc.

% Figure \ref{fig:histogram_SFR} shows the logarithmic SFR histograms for the major interactions and snapshots within 0.5 Gyr after the corresponding major interactions. There are no significant differences between the two samples; a K-S test between the two populations produces a p-value of 0.14. This analysis is insensitive to the distance criterion being used for interactions, with no significant difference seen when limiting the interaction distance to 50 kpc.
%with the non-major merger sample showing a slightly higher SFR than the merger sample.

We have shown that in our simulations galaxy interactions do not cause the majority of starbursts, so starbursts can occur in the absence of a major merger. However, we also see instances of major mergers that do result in starbursts. To see why this might be the case, we also looked at the influence of orbital parameters of companion galaxies. For example, prograde mergers are more likely to excite resonant \citep{D'Onghia2010ApJ...725..353D} interactions between the gas and stars in the host and the orbital frequency of the perturber, so perhaps prograde mergers more often produce starbursts  \citep{D'Onghia2010ApJ...725..353D}. However, most of the companion galaxies in the major interaction sample are not in strictly prograde or retrograde orbits because the central galaxies have not developed well-formed disks yet. The inclinations between the angular momentum vectors of the orbits of the companion galaxies and the angular momentum vectors of the central galaxies are shown in Figure \ref{fig:inclination_stacked} in Appendix \ref{app_b} for all simulations that involve major mergers. Thus we do not have any evidence that the orbital parameters have a significant effect on the star formation in the host galaxy. %Only in one case does the major merger have a prograde orbit -- the second merger in `m12w' -- which is also the latest-occurring merger in the sample at $z \sim 0.6$.

The detrended SFRs for `m12m' as shown in Figure \ref{fig:gas_star_dSFR_m12m} and similar figures for other runs (Figure \ref{fig:gas_star_dSFR_m12c}, and Figures \ref{fig:gas_star_dSFR_m12w} and \ref{fig:gas_star_dSFR_m12f} in Appendix \ref{app_c}) demonstrate that the SFR history across the simulations look similar regardless of the presence of major interaction events, transitioning from `bursty' to `steady' star-forming state during the redshift range in the analysis. 

As shown in Figure \ref{fig:SFR_specific_torque_m12m}, and Figures \ref{fig:SFR_specific_torque_m12b} and \ref{fig:SFR_specific_torque_m12w} in Appendix \ref{app_a}, the star formation history of the galaxies across cosmic time are not dramatically affected by major interactions. Even in the cases where major interactions are present and the torques are strongly correlated with the starbursts, the major interactions do not leave a significant imprint on the total star formation history of the galaxies. The only exception is `m12c', which has an abrupt jump in stellar mass right before the end of the second major merger, as shown in Figure \ref{fig:gas_star_dSFR_m12c} and appendix Figure \ref{fig:SFR_specific_torque_m12b}.

\section{Discussion}
\label{sec:discussion}

The above analysis of the `m12' simulations indicates that the occurrence of starbursts in Milky Way-mass galaxies are mostly independent of interactions of companion halos. Furthermore, if any mechanism from a flyby or merger enhances the star formation within the central galaxy, its overall effect is weak.

A difficulty in the methodology of this analysis comes from the need for the central galaxy and companion galaxy to be separable and independently traceable. However, as the secondary galaxy approaches within 10 kpc of the center of the main halo, the simulation particles are substantially mixed, preventing a clean separation between the two components, as would be expected during the merger process. Consequently, any starburst immediately after the merger of the systems would be missed by the correlation metrics used. However, other studies have found that the majority of the star formation enhancement occurs prior to coalescence of the two systems \citep[e.g.][]{Martin2017}, which we find to be consistent with this work, as none of the systems shows a starburst immediately after the completion of the merger.

% When calculating the correlation between star formation and torque due to satellite galaxies, we do not account for enhancements in the SFR arising from the fact that the two progenitor galaxies have merged into one, with higher stellar and (possibly) gas mass: when the two merging galaxies are identified as the same structure by Rockstar, we do not have a separate satellite galaxy, and therefore there is no torque to correlate with the SFR during that time period. Other studies have found that the majority of the SFR enhancement occurs before coalescence \citep{Martin2017}, if that holds true for the simulations in this study, it means that missing the remnant phase has a limited influence on the results. Additionally, in this paper we only aim to investigate the star formation enhancement effect of mergers, not to measure the total amount of star formation enhancement, which is worth studying in future work.

The analysis conducted in this study is motivated by the suggestion of a specific proposed physical mechanism of interaction-induced star formation enhancement: torque-induced nuclear inflow of gas. While the overall analysis aims to identify any correlation between the companion galaxy behavior and star formation in the central galaxy, the detailed correlation analysis between the torque induced by the secondary galaxy and the SFR of the main galaxy is designed to specifically identify any effect of this pathway on starbursts. Other mechanisms of central-companion galaxy interplay have been suggested as possible sources of star formation enhancement, including shocks and tidal compression \citep{Jog1992, Renaud2014, Renaud2019}; however, a specific analysis of these pathways is beyond the scope of this current work. However, we note that the analysis conducted in this work puts an upper envelope on the total star formation enhancement that could be produced by any form of interaction for Milky Way-mass galaxies independent of physical mechanism. 

% Discussion of Mechanisms
% In this study, we explored one physical mechanism in merger-induced star formation enhancement -- the torque-induced nuclear inflow. There are other processes that can play important roles in inducing starbursts during different stages of galaxy interactions, such as shocks and tidal compression \citep{Jog1992, Renaud2014, Renaud2019}. The inclusion of the other physical mechanisms in the analysis can be especially important in early interaction phases, as the gravitational torque is not as strong at high impact parameters, and they can also potentially be important in extended off-nuclear regions of star formation when the global gas inflow is not relevant. The current analysis is likely sufficient when it comes to the inner regions of advanced mergers, The inclusion of the other mechanisms in future work would allow for a better understanding of the merger-induced star formation during all stages of galaxy interaction.

\subsection{Comparison to Observations}
The major mergers in this study occur during the redshift range $z=1.2-3.6$, which we can compare to observational studies of galaxies in that same time range. 

\citet{Pearson2019} and \citet{Silva2018} both compare the star formation activity of merging and non-merging galaxies through the HST CANDELS survey: in both cases, they find an insignificant difference between the merger and non-merger star formation rates between galaxies up to a redshift of 4. In Figure \ref{fig:histogram_SFR}, we show the histogram of the star formation rates of each snapshot broken into major merger and non-major merger populations, showing no statistical difference between the populations, in excellent agreement with these observational works. 

A wider observational study by \citet{Ellison2008} using SDSS galaxies finds an enhancement in the star formation of galaxies with close ($<40$ kpc) interactions as compared to field galaxies, though the effect is most prominent at the closest radii and with mass ratios greater than 0.5. We note that the statistical elevation measured by \citet{Ellison2008}, up to a 70\% increase in the average SFR, would not constitute a ``starburst'' event for the purposes of this study. This result is overall consistent with the mild correlation we measure between the torque and the SFR for major interactions. 

Further, the starburst rates inferred from these simulations are largely consistent with observational studies. We infer a starburst rate of $3\%$ between $0<z<3.6$, with no starburst occurring after $z=0.4$ (Table \ref{tab:starburst_number}). About $15\%$ of the starbursts occur within 0.5 Gyr after major interactions, consistent with the results from \citet{Muratov2015MNRAS.454.2691M} and \citet{Angles2017MNRAS.470.4698A}, which suggest bursts followed by gas recollapsing onto the central halo to cause subsequent bursts.

\citet{Bergvall2016} find a starburst rate of $1\%$ for $10^{9} \rm M_{\rm \odot}<\rm M_{\rm gas+stars}<10^{11.5} \rm M_{\rm \odot}$ galaxies in the local universe, albeit with a different definition of starburst corresponding to events raising the star formation to three times the lifetime average SFR of the galaxy, versus our threshold of four times the local average SFR. \citet{Rodighiero2011} find a starburst rate of $2-3\%$ for $10^{10} \rm M_{\rm \odot}<\rm M_{\rm *}<10^{12} \rm M_{\rm \odot}$ galaxies at $1.5<z<2.5$ using a starburst threshold of four times the typical SFR as inferred from the galaxy main sequence. Overall, the FIRE simulations are consistent with the observed statistics of starburst events from observations.

\subsection{Comparison with Previous Simulations}

The FIRE simulations provide a high-resolution cosmological view of the evolution of Milky Way-mass galaxies, taking into account their full assembly and environmental history. This provides us with a unique comparison to other simulations that focus on isolated systems and other cosmological simulations. We do note that the burstiness of the star formation in the FIRE simulations is due to their high-resolution of galaxy ISM, which have no real analogue in other simulation work discussed here \citep{Wetzel2023ApJS..265...44W}.

This work is consistent with the analyses from other cosmological simulations, such as \citet{Martin2017}, using the Horizon-AGN simulations. They also find that the merger contribution to stellar mass growth is small at all redshifts, and not a dominant driver across the life of the galaxy; see also \citet{Keres2005MNRAS.363....2K} for the same conclusion.

Further, the current analysis provides useful context to simulations of merging galaxies in isolated volumes: \citet{Perret2014} modeled major and minor interactions, using idealized galaxies with high gas fractions at $1 < z < 2$, finding no enhancement of overall star formation. Additionally, idealized simulations provide opportunities to tweak experimental parameters to measure their effect, for instance, \citet{Fensch2017} infer the efficiency of high-redshift mergers to be significantly lower than those at low-redshift, while \citet{Scudder2015} find that galaxies with high gas fractions have higher baseline SFRs and weaker star formation enhancements than lower gas fraction galaxies (\citet{Robertson2006ApJ...645..986R} find a similar relationship between SFRs during the first pericentric passage and final coalescence). \citet{Moreno2021} use high-resolution, isolated simulation volumes of major mergers to identify a global $\sim$35\% enhancement of the star formation rate in the primary galaxy due to a merger.  

However, with the cosmological perspective provided by the FIRE simulations, we find that low-redshift interactions with Milky Way-mass galaxies are rare in comparison to higher redshifts, and the star formation history and gas fractions of these galaxies have little variation for samples at the same redshift. Consequently, the cosmological perspective allows us to exclude physical mechanisms that are unlikely to proceed due to the environmental and merger history of such halos.

One finding from \citet{Cenci2024MNRAS.527.7871C}, which also used cosmological zoom-in simulations with FIRE-2 physics, is that major mergers result in the largest difference in the proportion of interacting starburst and interacting non-starburst galaxies. It is consistent with the mild correlation in minor and mini interactions between torque and detrended SFR in this paper.

\section{Conclusions}
\label{sec:conclusions}

We use six cosmological zoom simulations of Milky Way-mass star-forming galaxies from the FIRE-2 simulation suite to investigate whether interactions with flyby or merging halos may induce starbursts in such galaxies. We trace interacting halos around the central galaxy for each simulation suite and measure the torque of these halos on the central galaxy's gas content. We measure the correlation between the torque and star formation during all merger and flyby events from $z=3.6$ to $z=0$, separating the interactions into major, minor, and mini interactions based on the stellar mass ratio of the galaxy pairs. We similarly measure the baseline torque-star formation correlation using all other snapshots. Our main conclusions are as follows:

(i) For major interactions, there is a positive and statistically significant correlation between the torque from nearby galaxies on the gas of the central galaxy and the detrended SFR. The correlation results in a Spearman coefficient of 0.41 with a p-value of $4.1 \times 10^{-12}$ (Table \ref{tab:correlation_combined})

(ii) There is no one particular pathway where interactions lead to starbursts in the central galaxy: Some major interactions cause starbursts, but most starbursts are not caused by galaxy interactions.

(iii) For minor interactions, mini interactions, and all other snapshots in the sample, there is no statistically significant correlation between torque and star formation (Table \ref{tab:correlation_combined}).

% (iv) The minimum distance of nearby halos to the center of Milky Way-mass galaxies increases with time, with few close interactions occurring for $z < 1.2$ (Figure \ref{fig:min_distance_m12m}).
% The flyby galaxies formed later has lower concentration, and as they merge into the main galaxy, as shown in Figure \ref{fig:min_distance_m12m}, they get disrupted at a larger distance to the centre of the main halo compared to the satellites at higher redshift. 

(iv) The transition from `bursty' to `steady' star-formation state appears to be independent of the interaction history of the galaxies.

(v) Most halo interactions do not leave a significant imprint on the overall trend of the star formation history of Milky Way-mass galaxies.

This work argues for a paradigm where the star formation processes of Milky Way-mass galaxies is not strongly dependent on the physics of interactions (flybys or mergers) but rather due to cosmological environments that control the overall accretion rates of the halos. In this picture, Milky Way-mass galaxies undergo most of their major interaction activities at early times, undergo only a handful of major mergers at most, and have star formation histories, in particular, burstiness, driven primarily by internal dynamics. 

\section*{Acknowledgements}

%We thank Jorge Moreno for his valuable comments which have improved this paper. 
NWM acknowledges the support of the Natural
Sciences and Engineering Research Council of Canada (NSERC; RGPIN-2023-04901). This work was performed in part at the Aspen Center for Physics, which is supported by National Science Foundation grant PHY-2210452.
DK was supported by NSF grant AST-2108324. AW received support from NSF, via CAREER award AST-2045928 and grant AST-2107772, and HST grant GO-16273 from STScI. CAFG was supported by NSF through grants AST-2108230 and AST-2307327; by NASA through grant 21-ATP21-0036; and by STScI through grant JWST-AR-03252.001-A. Support for PFH was provided by NSF Research Grants 1911233, 20009234, 2108318, NSF CAREER grant 1455342, NASA grants 80NSSC18K0562, HST-AR-15800. The simulations presented here used computational resources granted by the Extreme Science and Engineering Discovery Environment (XSEDE), which is supported by National Science Foundation grant no. OCI-1053575, specifically allocation TG-AST120025 and resources provided by PRAC NSF.1713353 supported by the NSF; Frontera allocations AST21010 and AST20016, supported by the NSF and TACC; Blue Waters, supported by the NSF; Pleiades, via the NASA HEC program through the NAS Division at Ames Research Center. The analysis of the FIRE simulation data was run on the CITA computing cluster ``Sunnyvale''. Computations were performed on the Niagara supercomputer \citep{10.1145/3332186.3332195} at the SciNet HPC Consortium \citep{Loken_2010}. SciNet is funded by Innovation, Science and Economic Development Canada; the Digital Research Alliance of Canada; the Ontario Research Fund: Research Excellence; and the University of Toronto. The data used in this work were, in part, hosted on facilities supported by the Scientific Computing Core at the Flatiron Institute, a division of the Simons Foundation. 
 
The research made use of the following software: SCIPY \footnote{http://scipy.org}, NUMPY \footnote{http://numpy.org} \citep{Walt2011}, MATPLOTLIB \footnote{http://matplotlib.org} \citep{Hunter2007}, and YT \footnote{http://yt-project.org} \citep{Turk2011}.

%%%%%%%%%%%%%%%%%%%%%%%%%%%%%%%%%%%%%%%%%%%%%%%%%%
\section*{Data Availability}

A public version of the GIZMO code is available at http://www.tapir.caltech.edu/~phopkins/Site/GIZMO.html. FIRE-2 simulations are publicly available \citep{Wetzel2023ApJS..265...44W} at http://flathub.flatironinstitute.org/fire. Additional data, including initial conditions and derived data products, are available at https://fire.northwestern.edu/data/.



\bibliographystyle{mnras}
\bibliography{example} % if your bibtex file is called example.bib

%\begin{thebibliography}{99}
%\end{thebibliography}

%%%%%%%%%%%%%%%%%%%%%%%%%%%%%%%%%%%%%%%%%%%%%%%%%%

%%%%%%%%%%%%%%%%% APPENDICES %%%%%%%%%%%%%%%%%%%%%

\appendix

\eject

% Start correlation plots

%%%%%%%%%%%%%%%%%%%%%%%%%%%%%%%%%%%%%%%%%%%%%%%%%%%%%%%%%%%%

\section{Torque and Detrended SFR Figures}
\label{app_a}

\begin{figure*}
% figure from /GIZMOUtils.jl/Tracing/plotting/Specific_Torque_SFR_all_halos.ipynb
	\includegraphics[width=\textwidth]{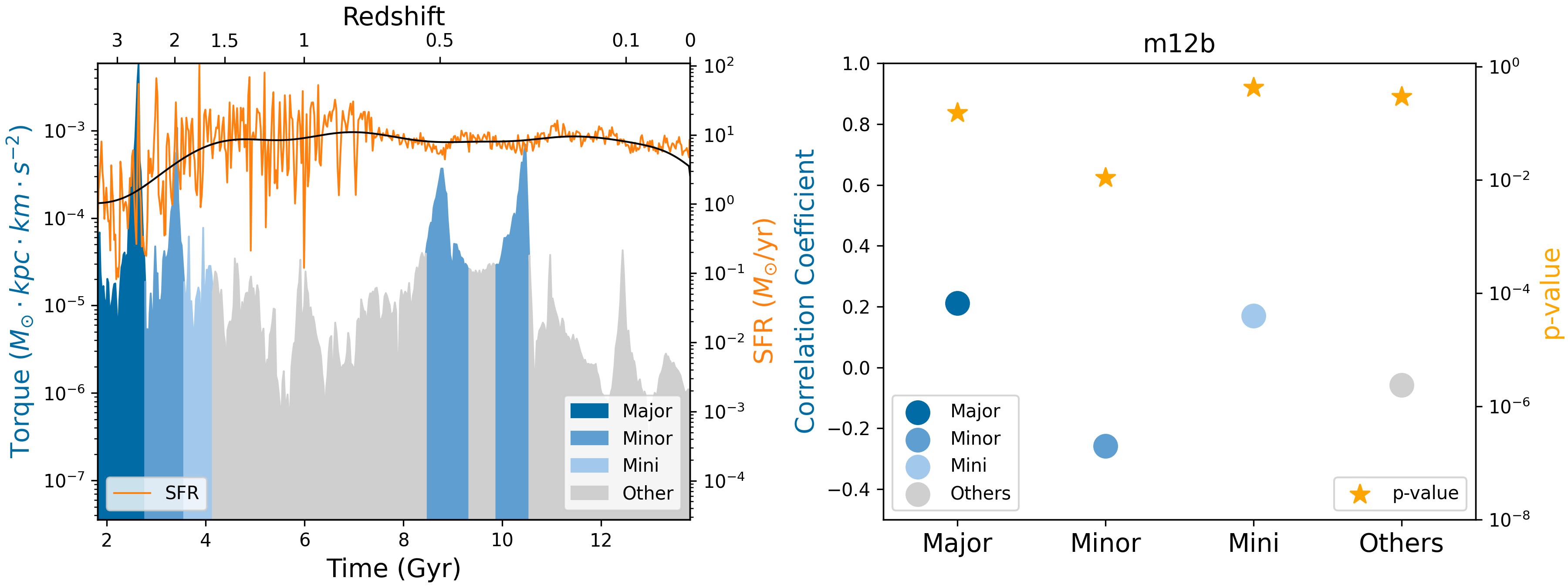}
    \includegraphics[width=\textwidth]{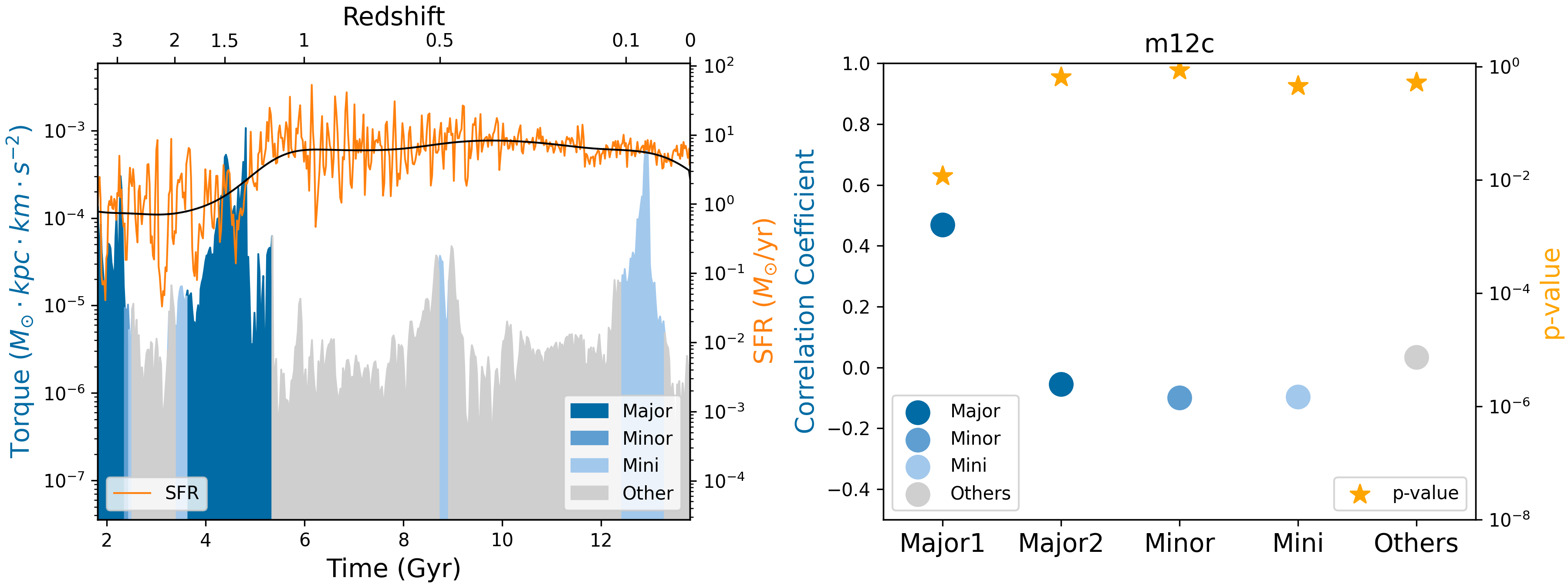}
    \includegraphics[width=\textwidth]{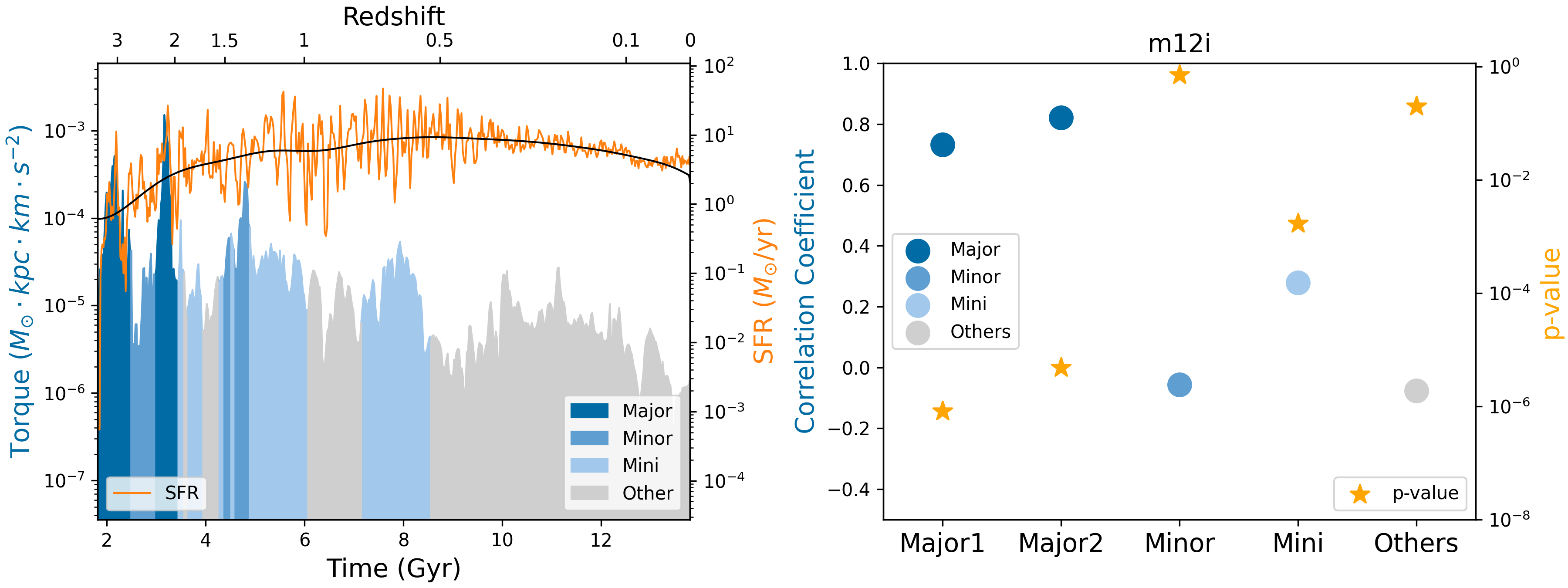}
    \caption{Analagous to Figure \ref{fig:SFR_specific_torque_m12m}, but for the simulated galaxies `m12b', `m12c', and `m12i'.} %the correlation coefficients are calculated for the 5 equal snapshot bins between snapshot 100 (z=3.4) and snapshot 600 (z=0). }
    \label{fig:SFR_specific_torque_m12b}
\end{figure*}

\begin{figure*}
\centering
% figure from /GIZMOUtils.jl/Tracing/plotting/Specific_Torque_SFR_all_halos.ipynb
 \includegraphics[width=\textwidth]{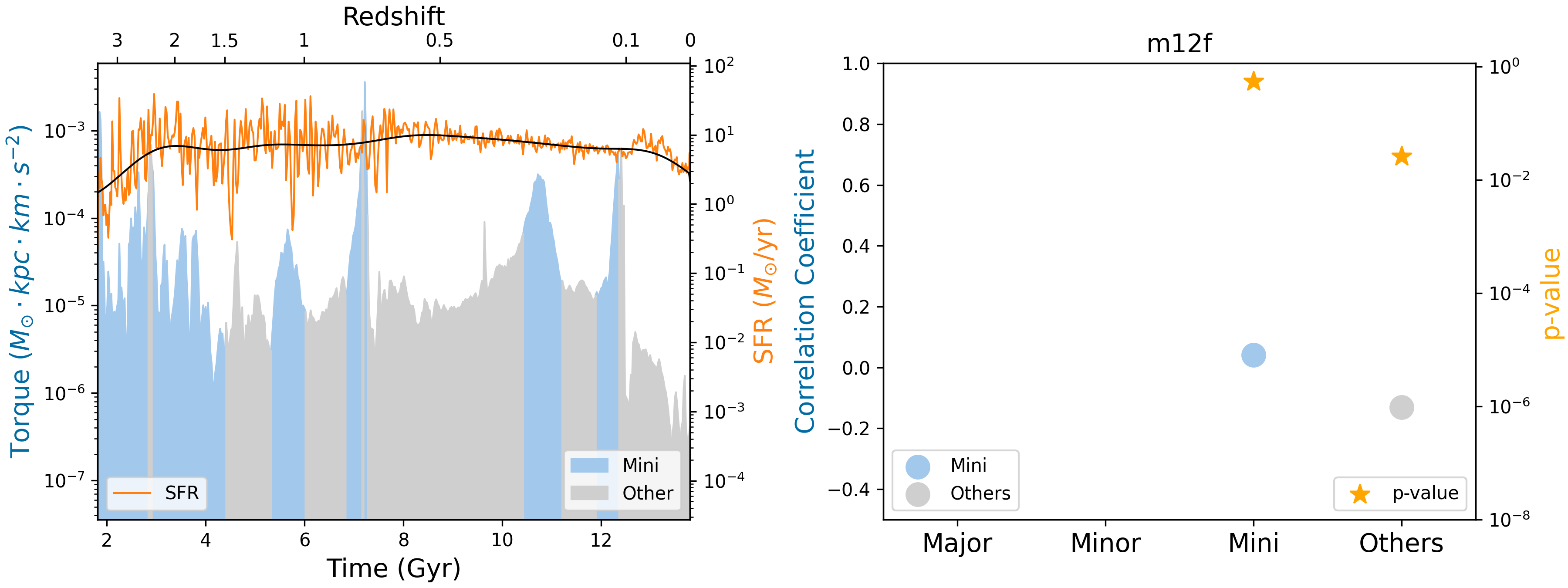}
 \includegraphics[width=\textwidth]{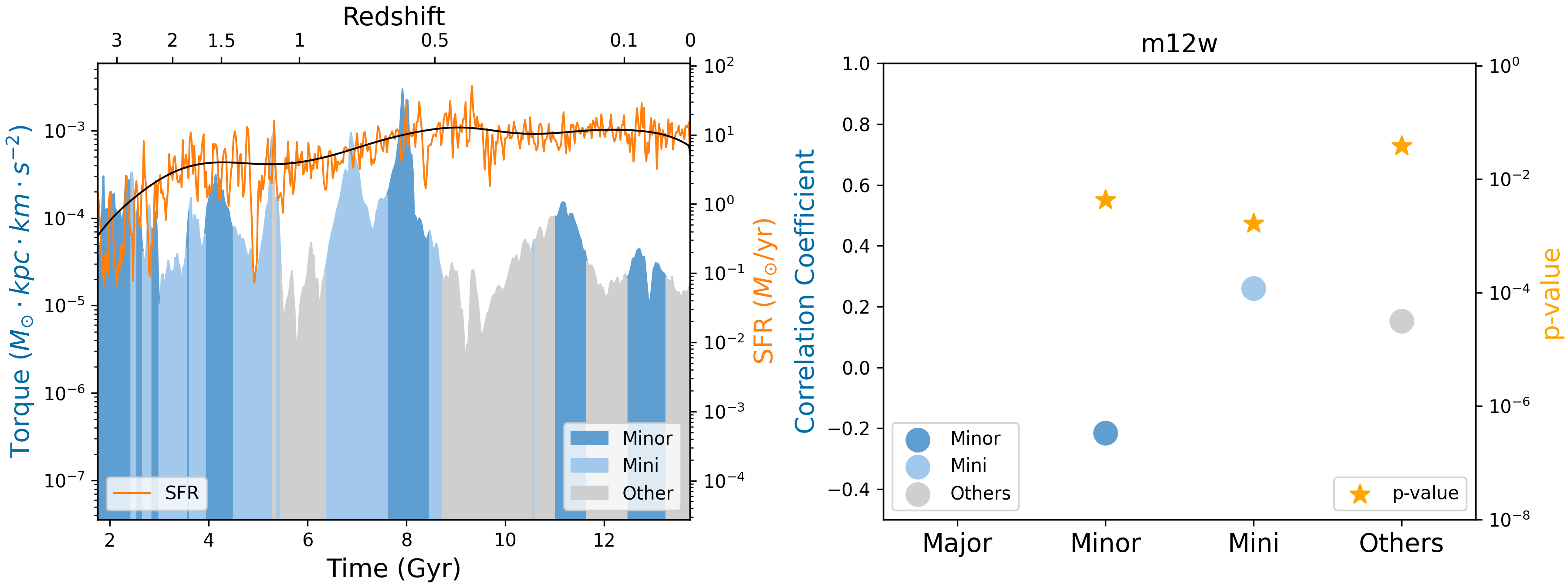}
    \caption{Analagous to Figure \ref{fig:SFR_specific_torque_m12m}, but for the simulated galaxies `m12f' and `m12w'.}
    \label{fig:SFR_specific_torque_m12w}
\end{figure*}

Figure \ref{fig:SFR_specific_torque_m12m} in the main text shows the torque and the SFR along with the correlation coefficients and associated p-values for different interaction groups for the simulated galaxy `m12m'. Figures \ref{fig:SFR_specific_torque_m12b} and \ref{fig:SFR_specific_torque_m12w} in this appendix show the same relationship for the other simulated galaxies in this study. The values of the correlation coefficients are shown in Table \ref{tab:correlation_torque_spearman_detrended}.

%\begin{figure*}
% figure from /GIZMOUtils.jl/Tracing/plotting/Specific_Torque_SFR_all_halos.ipynb
%	\includegraphics[width=\textwidth]{bar_m12c.png}
%    \caption{Same as Figure \ref{fig:SFR_specific_torque_m12b}, but for the simulated galaxy `m12c'.}
%    \label{fig:SFR_specific_torque_m12c}
%\end{figure*}

%\begin{figure*}
% figure from /GIZMOUtils.jl/Tracing/plotting/Specific_Torque_SFR_all_halos.ipynb
%	\includegraphics[width=\textwidth]{bar_m12r.png}
%    \caption{Same as Figure \ref{fig:SFR_specific_torque_m12b}, but for the simulated galaxy `m12r'.}
%    \label{fig:SFR_specific_torque_m12r}
%\end{figure*}

%\begin{figure*}
% figure from /GIZMOUtils.jl/Tracing/plotting/Specific_Torque_SFR_all_halos.ipynb
%	\includegraphics[width=\textwidth]{bar_m12f.png}
% \includegraphics[width=\textwidth]{bar_m12w.png}
%    \caption{Same as Figure \ref{fig:SFR_specific_torque_m12b}, but for the simulated galaxy `m12f' and `m12w'.}
 %   \label{fig:SFR_specific_torque_m12f}
%\end{figure*}

%\begin{figure*}
% figure from /GIZMOUtils.jl/Tracing/plotting/Specific_Torque_SFR_all_halos.ipynb
%	\includegraphics[width=\textwidth]{bar_m12w.png}
%    \caption{Same as Figure \ref{fig:SFR_specific_torque_m12b}, but for the simulated galaxy `m12w'.}
%    \label{fig:SFR_specific_torque_m12w}
%\end{figure*}

\section{Inclination Figures}
\label{app_b}

\begin{figure*}
\centering
\includegraphics[width=\textwidth,height=\textheight,keepaspectratio]{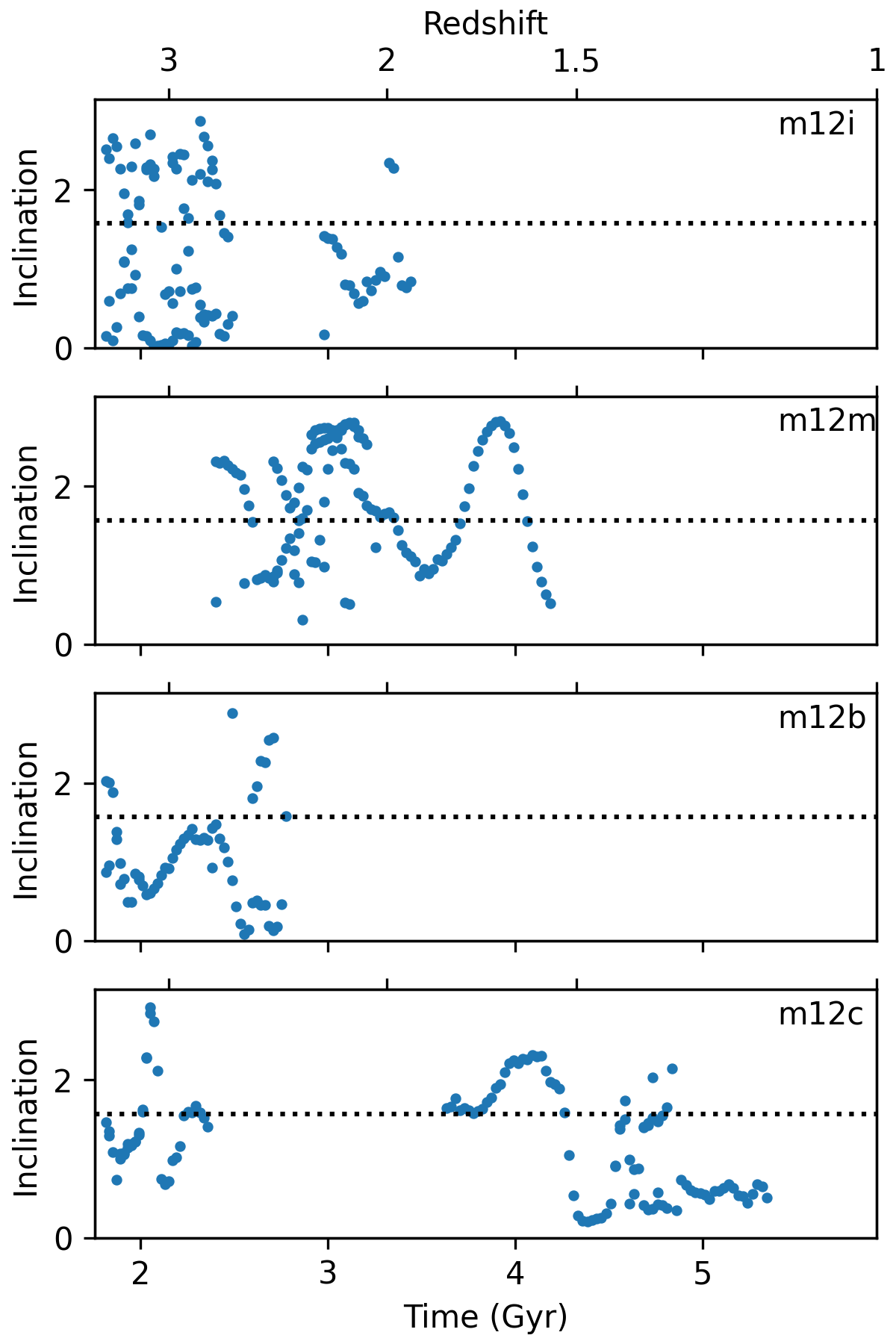}
    \caption{The angle (in radians) between the angular momentum vector of the main galaxies and the angular momentum vector of the companion galaxies for major interactions as a function of cosmic time for the simulation runs with major interactions.}
    \label{fig:inclination_stacked}
\end{figure*}

To study the impact of retrograde and prograde orbits on the correlation between the torque and the SFR, we calculated the angle between the angular momentum vector of the main galaxy and the angular momentum vector of the companion galaxy for major interactions as a function of cosmic time for all snapshots undergoing major interactions, shown in Figure \ref{fig:inclination_stacked}.

%\begin{figure}
%\includegraphics[width=\columnwidth,height=\textheight,keepaspectratio]{inclination_m12i.png}
%    \caption{The angle (in radians) between the angular momentum vector of the main galaxy and the angular momentum vector of the satellite galaxy for major mergers as a function of cosmic time for the simulation run `m12i'.}
%    \label{fig:inclination_m12i}
%\end{figure}

%\begin{figure}
%\includegraphics[width=\columnwidth,height=\textheight,keepaspectratio]{inclination_m12c.png}
%    \caption{Same as Figure \ref{fig:inclination_m12i}, but for the simulated galaxy `m12c'.}
%    \label{fig:inclination_m12c}
%\end{figure}

%\begin{figure}
%\includegraphics[width=\columnwidth,height=\textheight,keepaspectratio]{inclination_m12b.png}
%    \caption{Same as Figure \ref{fig:inclination_m12i}, but for the simulated galaxy `m12b'.}
%    \label{fig:inclination_m12b}
%\end{figure}

%\begin{figure}
%\includegraphics[width=\columnwidth,height=\textheight,keepaspectratio]{inclination_m12m.png}
%    \caption{Same as Figure \ref{fig:inclination_m12i}, but for the simulated galaxy `m12m'.}
%    \label{fig:inclination_m12m}
%\end{figure}

\section{Gas Mass, Stellar Mass, And Detrended SFR Figures}
\label{app_c}

This section contains figures of gas mass, stellar mass, and the detrended SFR as a function of cosmic time; the plots for `m12m' (Figure \ref{fig:gas_star_dSFR_m12m}) and `m12c' (Figure \ref{fig:gas_star_dSFR_m12c}) are included in the main text, and the same plots for the other simulated central galaxies are shown in this section. All of the simulated galaxies in the sample have mass and star formation histories that are qualitatively similar. 

\begin{figure}
\centering
\includegraphics[width=\columnwidth,height=\textheight,keepaspectratio]{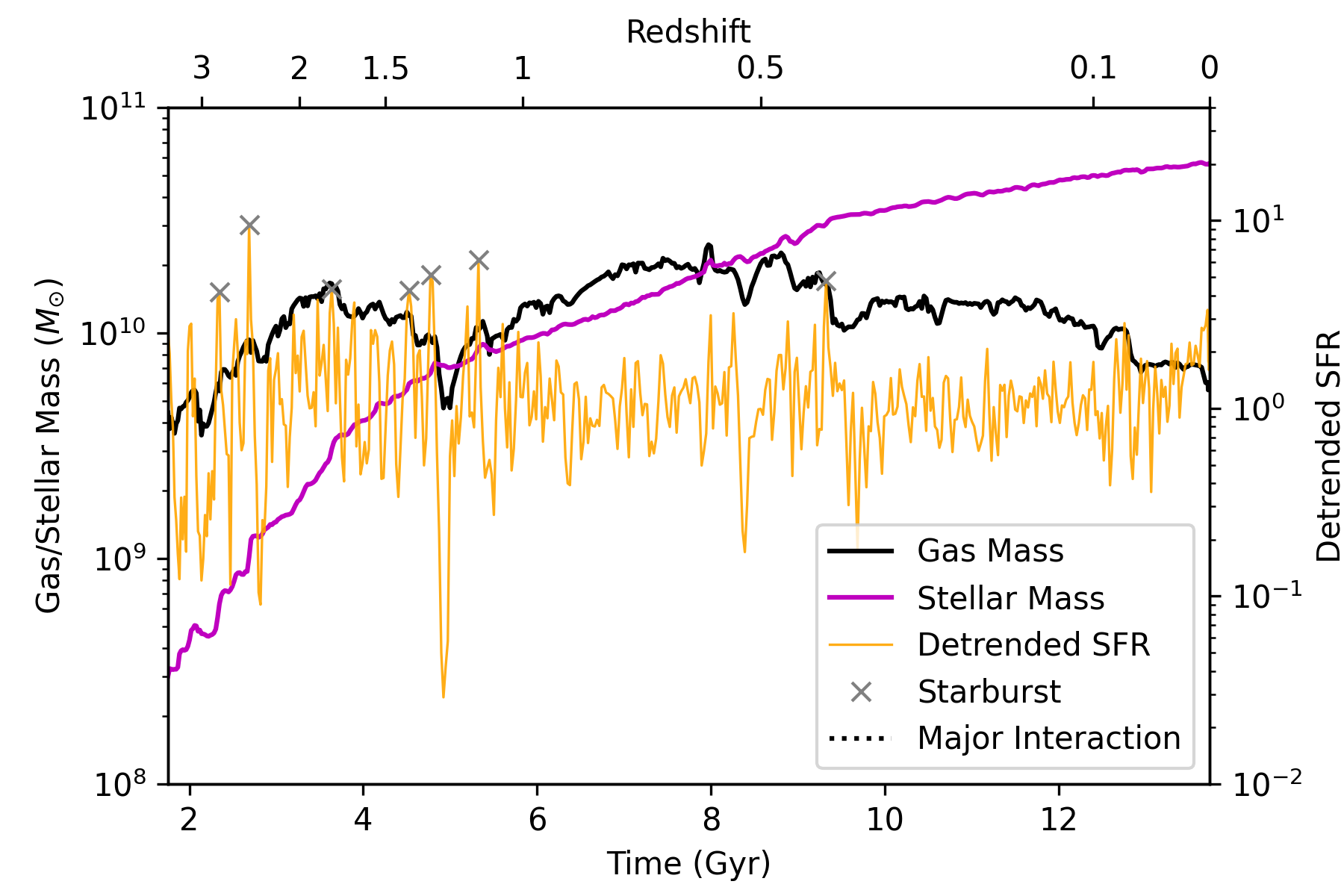}
\includegraphics[width=\columnwidth,height=\textheight,keepaspectratio]{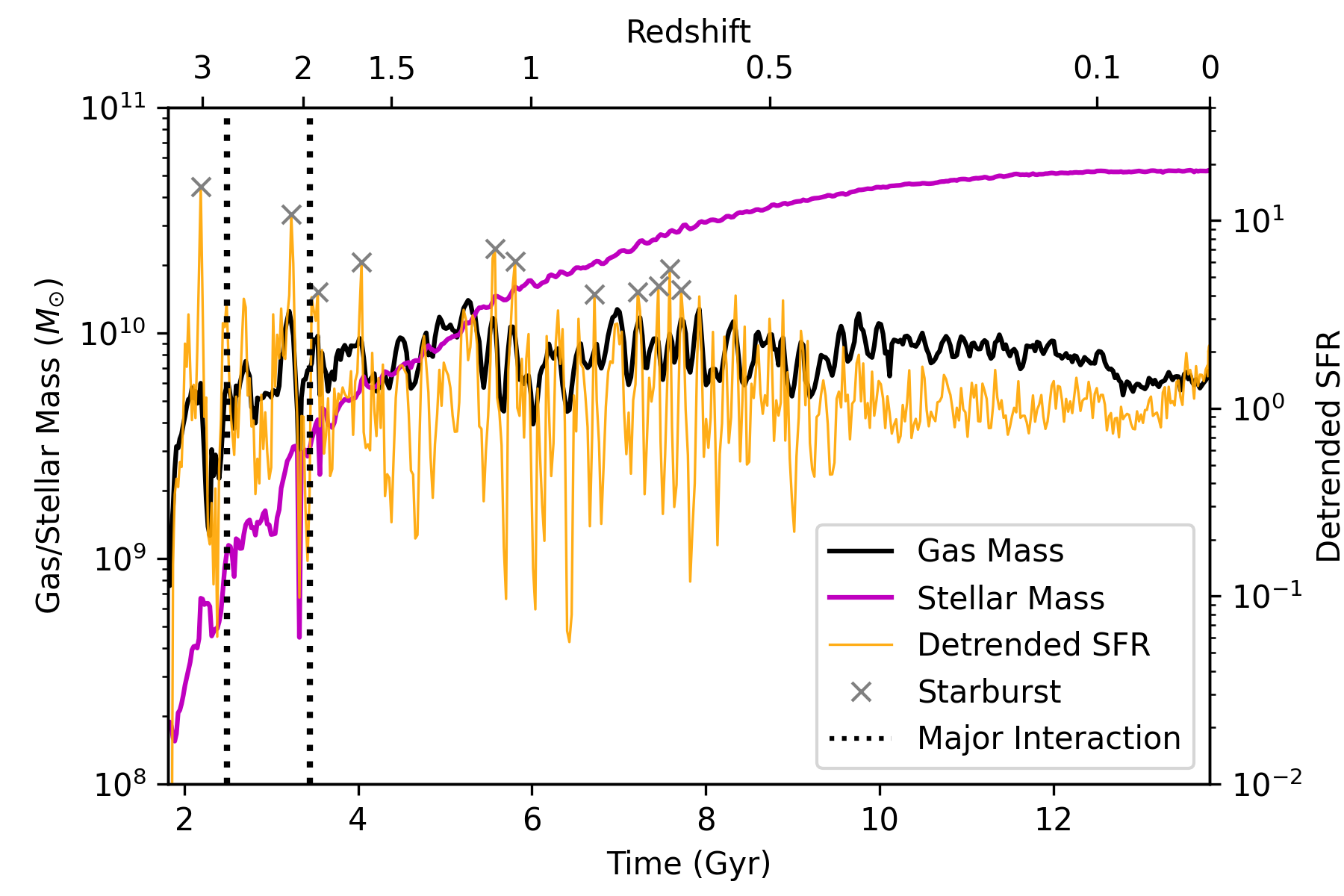}
    \caption{Analogous to Figure \ref{fig:gas_star_dSFR_m12m}, but for the simulated galaxy `m12w' and `m12i'.}
    \label{fig:gas_star_dSFR_m12w}
\end{figure}

%\begin{figure}
%\includegraphics[width=\columnwidth,height=\textheight,keepaspectratio]{gas_star_dSFR_m12i.png}
%    \caption{Same as Figure \ref{fig:gas_star_dSFR_m12w}, but for the simulated galaxy `m12i'.}
%    \label{fig:gas_star_dSFR_m12i}
%\end{figure}

%\begin{figure}
%\includegraphics[width=\columnwidth,height=\textheight,keepaspectratio]{gas_star_dSFR_m12c.png}
%    \caption{Same as Figure \ref{fig:gas_star_dSFR_m12w}, but for the simulated galaxy `m12w'.}
%    \label{fig:gas_star_dSFR_m12c}
%\end{figure}

\begin{figure}
\centering
\includegraphics[width=\columnwidth,height=\textheight,keepaspectratio]{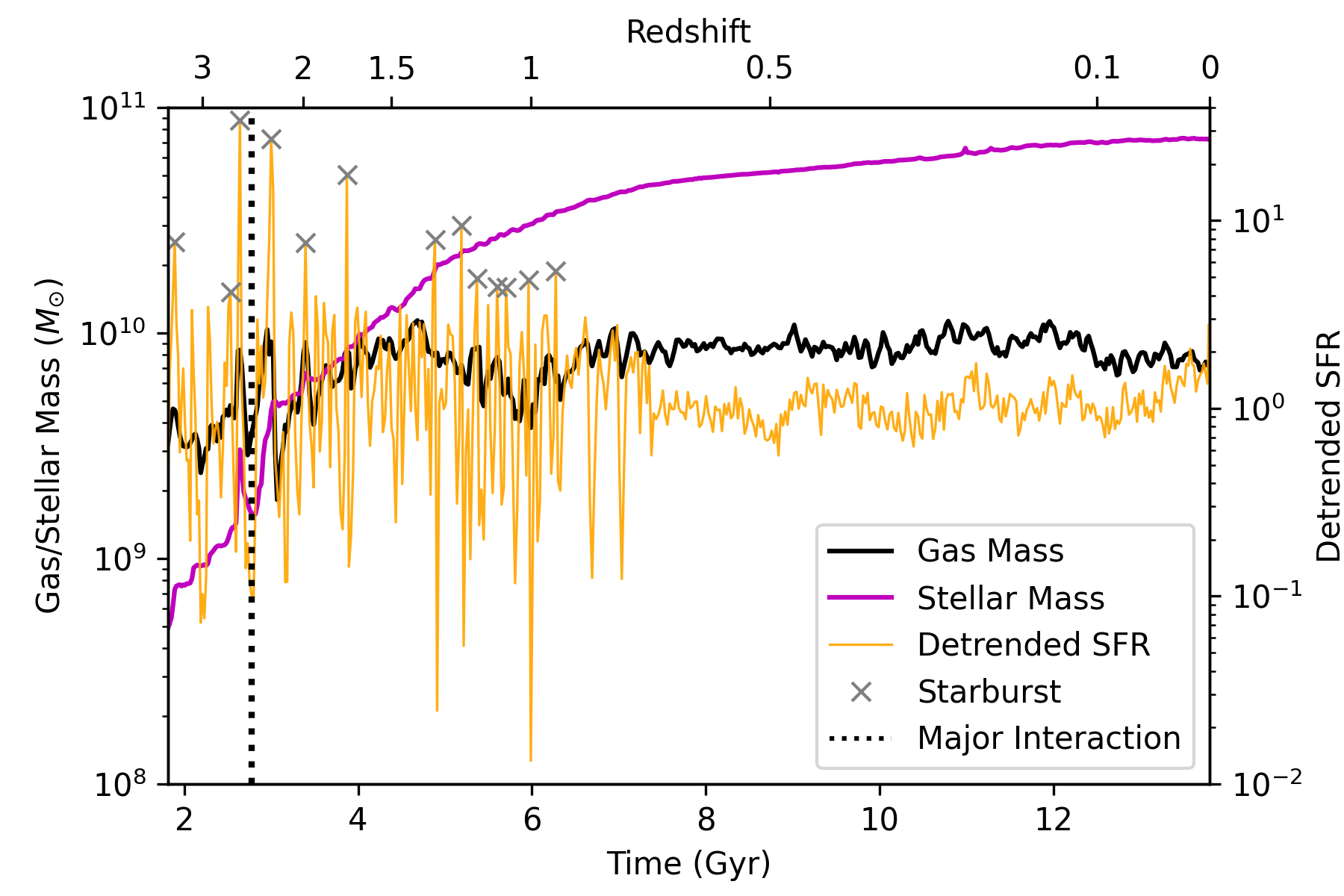}
\includegraphics[width=\columnwidth,height=\textheight,keepaspectratio]{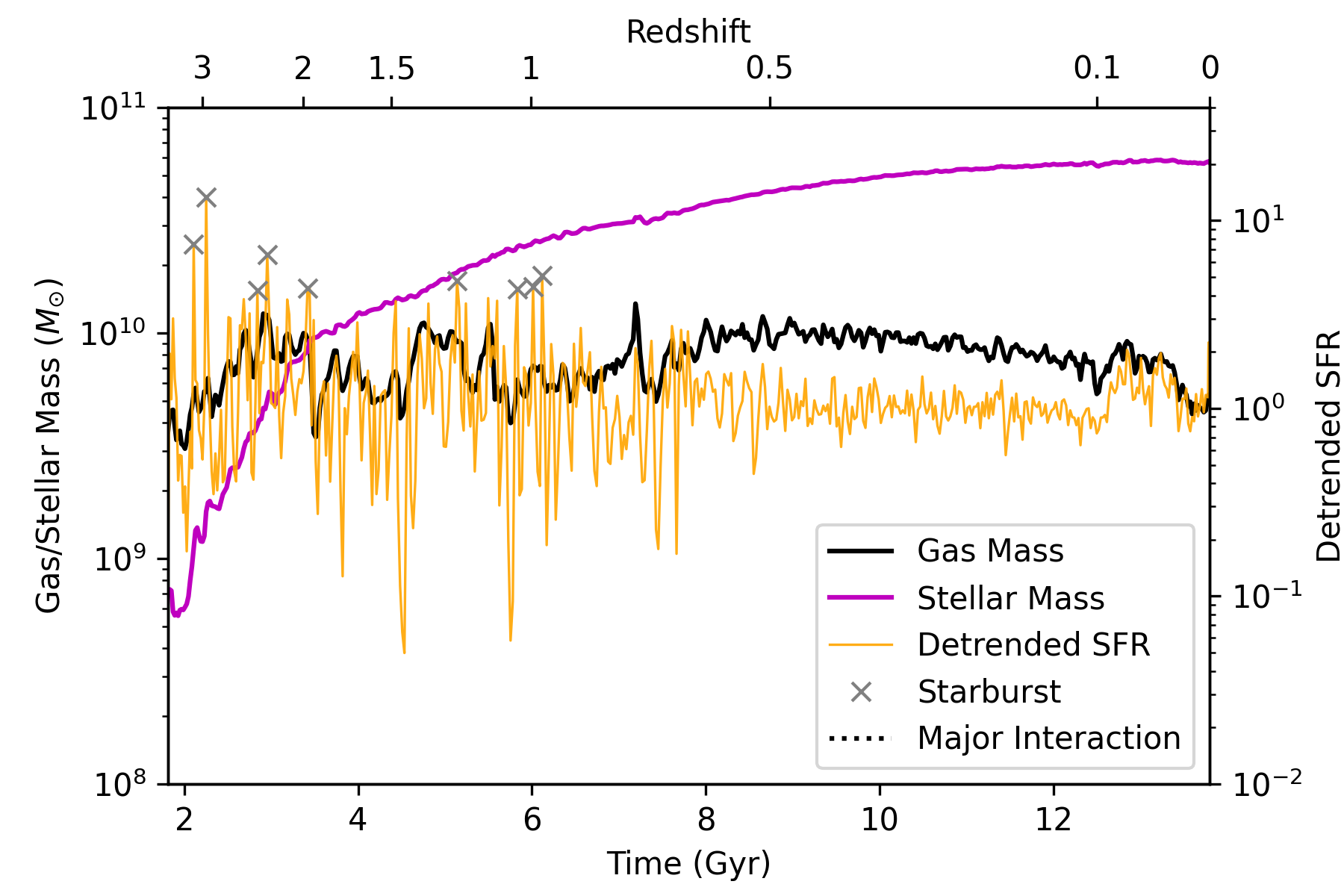}
    \caption{Analogous to Figure \ref{fig:gas_star_dSFR_m12m}, but for the simulated galaxy `m12b' and `m12f'.}
    \label{fig:gas_star_dSFR_m12f}
\end{figure}

%\begin{figure}
%\includegraphics[width=\columnwidth,height=\textheight,keepaspectratio]{gas_star_dSFR_m12f.png}
%    \caption{Same as Figure \ref{fig:gas_star_dSFR_m12w}, but for the simulated galaxy `m12f'.}
%    \label{fig:gas_star_dSFR_m12f}
%\end{figure}


\end{CJK*}
\end{document}